\documentclass[10pt,conference]{IEEEtran}

\IEEEoverridecommandlockouts
\usepackage{cite}
\usepackage{amsmath,amssymb,amsfonts}
\usepackage[linesnumbered,ruled,vlined,lined]{algorithm2e}

\SetCommentSty{mycommfont}
\usepackage{graphicx}
\usepackage{textcomp}
\usepackage{xcolor}
\usepackage[hyphens]{url}
\usepackage{fancyhdr}
\usepackage{hyperref}
\usepackage[caption=false]{subfig}

\begin{document}

% Ensure letter paper
\pdfpagewidth=8.5in
\pdfpageheight=11in

%%%%%%%%%%%---SETME-----%%%%%%%%%%%%%
\newcommand{\iscasubmissionnumber}{141}
%%%%%%%%%%%%%%%%%%%%%%%%%%%%%%%%%%%%

\pagenumbering{arabic}

%%%%%%%%%%%---SETME-----%%%%%%%%%%%%%
\title{TTP: A Hardware-Efficient Design for Precise Prefetching in Ray Tracing\thanks{†Both authors contributed equally to this work.}}
\author{\IEEEauthorblockN{Yavuz Selim Tozlu\textsuperscript{†}}
\IEEEauthorblockA{\textit{Electrical and Computer Engineering} \\
\textit{North Carolina State University}\\
Raleigh, USA \\
ystozlu@ncsu.edu}
\and
\IEEEauthorblockN{Anshul Naithani\textsuperscript{†}}
\IEEEauthorblockA{\textit{Electrical and Computer Engineering} \\
\textit{North Carolina State University}\\
Raleigh, USA \\
anaitha2@ncsu.edu}
\and
\IEEEauthorblockN{Huiyang Zhou}
\IEEEauthorblockA{\textit{Electrical and Computer Engineering} \\
\textit{North Carolina State University}\\
Raleigh, USA \\
hzhou@ncsu.edu}
}
%%%%%%%%%%%%%%%%%%%%%%%%%%%%%%%%%%%%

\maketitle
\thispagestyle{plain}
\pagestyle{plain}

\begin{abstract}
Ray tracing (RT) is a 3D graphics technique that offers highly realistic visuals. 
It is becoming prominent and accessible as GPU vendors have integrated dedicated ray tracing acceleration hardware. 
However, tracing millions of rays through 3D scenes consisting of high numbers of triangles in real time is challenging and requires expensive hardware. 
The main bottleneck in RT workloads is the expensive Bounding Volume Hierarchy (BVH) traversal task, which is a large tree structure that encodes the 3D scene.
BVH traversal is a memory-bound problem, as the GPU threads spend most of their time reading tree node data from memory.

In this work, we attack the memory latency bottleneck of ray tracing through prefetching.
We propose a novel hardware prefetcher, named Tree Traversal Prefetcher (TTP), for ray tracing. 
The main idea is to leverage the existing tree traversal stack in the RT units for highly accurate prefetching. 
In particular, TTP prefetches nodes using the addresses already available on the hardware traversal stacks of each thread.
For DFS (Depth-first search) based traversal, prefetches are generated when nodes are being popped consecutively from the traversal stack, potentially corresponding to upward traversal through the tree. 

We evaluate TTP on a cycle-level simulator, Vulkan-sim 2.0, and show that it achieves 1.48x speedup on average (up to 1.89x) compared to the baseline, with nearly negligible hardware overhead. 
TTP achieves 98.92\% average L1 accuracy, which is the ratio of the prefetched blocks being actually referenced by demand loads.
% The prefetching efficiency, defined as the ratio of prefetch requests that miss in L1, is 58.56\%.
%If the prefetches hitting in the cache or MSHRs are considered useless, the average accuracy is 54.84\% average.
The coverage, computed as the ratio of L1 miss reduction over baseline L1 misses, is 31.54\%, correlating well with the achieved speedup.
\end{abstract}

\section{Introduction} \label{introsec}

Ray tracing is a modern 3D rendering technique that generates highly realistic graphics for both real-time applications like video games and offline rendering such as animations\cite{unreal}\cite{frostbite}\cite{cars}. 
Unlike rasterization, ray tracing accurately models the lighting effects in a scene, providing life-like graphics\cite{nvidia_diff}. 
The challenge, however, is that ray tracing incurs high computational costs. Recently, hardware vendors have introduced specialized units, RT units/cores, in GPUs to accelerate ray tracing\cite{intel_rt}\cite{ada_arch}\cite{ampere_arch}\cite{burgess_rtx_2020}, prompting developers to adopt advanced ray tracing algorithms, thereby advancing real-time rendering\cite{frostbite}\cite{noauthor_unity_nodate}. 
Beyond computer graphics, RT units have been exploited for more general purpose applications.\cite{barnes_extending_2024}\cite{feng_heliostat_2025}\cite{geng_librts_2025}\cite{ha_generalizing_2024}\cite{mandarapu_arkade_2024}\cite{zhang_rtspmspm_2025}.

Despite advances in ray tracing hardware, rendering complex 3D scenes in real time remains challenging. 
First, each frame requires tracing millions of rays to generate a high resolution picture. 
Second, although different rays traverse the 3D scene independently, thereby being massively parallel, each ray follows its own traversal path to bounce through objects to find intersections, leading to unpredictable and divergent behavior. 
Thirdly, during the traversal of the objects in a scene, each ray needs to access a large data set, i.e., the Bounding Volume Hierarchy (BVH). 
As a result, on-chip caches are usually not sufficient, and the memory wall is exposed as a key performance bottleneck.

BVH traversal makes up the bulk of ray tracing. %TODO: add references.
A 3D scene is organized as a hierarchy of bounding boxes, which is stored as a tree in memory. 
The size of BVH trees can be in the order of gigabytes, depending on the scene complexity, thereby stressing the on-chip cache hierarchy.
During traversal, an RT unit fetches tree nodes from memory, tests rays for intersections with boxes, and descends through the tree to find the closest hit primitive for each ray. 
As intersection tests and coordinate transformations are carried out in fast, fixed-function hardware, memory bottleneck is exposed as the primary challenge. 

\begin{figure}[]
\centerline{\includegraphics[width=0.48\textwidth]{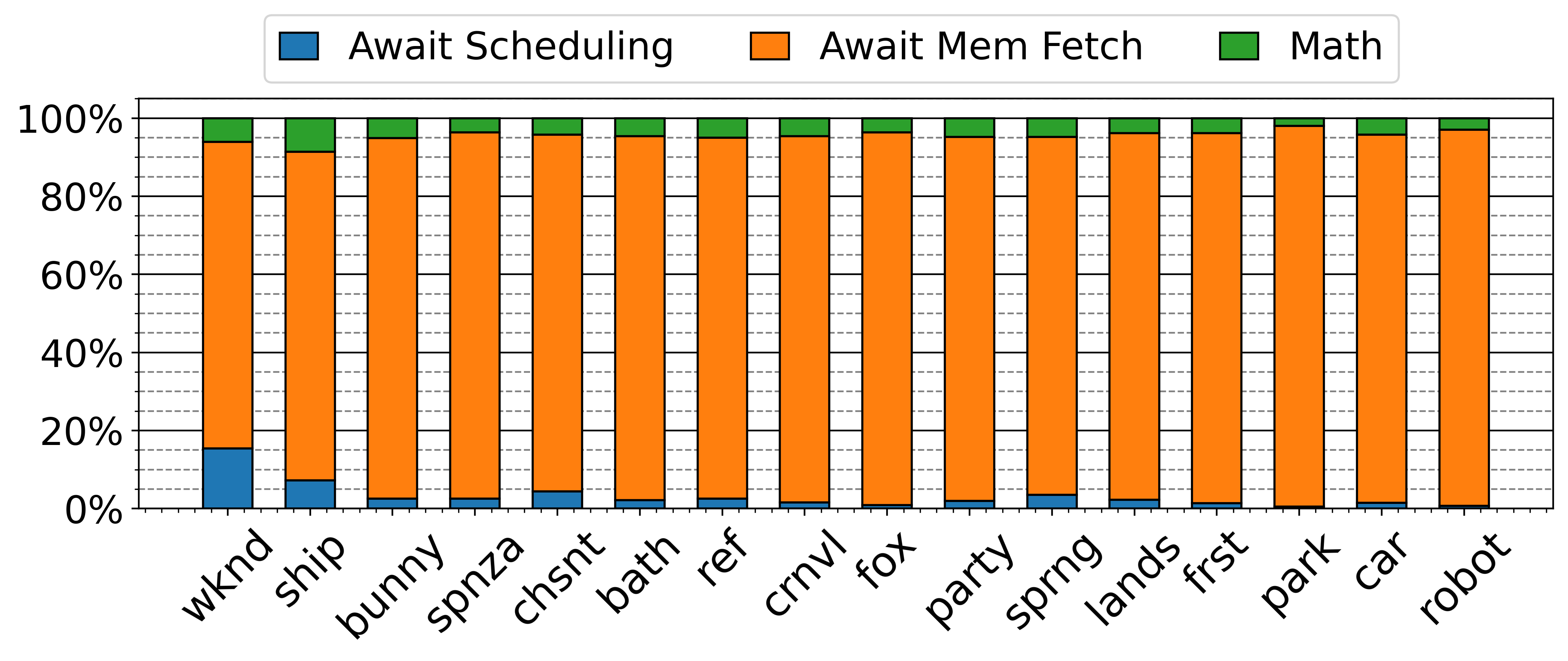}}
\caption{Thread status distribution in an RT unit. Threads may be waiting for scheduling, waiting on a memory fetch, or performing math operations such as intersection tests or coordinate transformations. 128x128 resolution path tracing, 1 sample per pixel.}
\label{rtcyclesdist}
\end{figure}

\begin{figure}[]
    \centering
        \subfloat[Average DRAM bandwidth utilization.]{
            \includegraphics[width=0.48\textwidth]{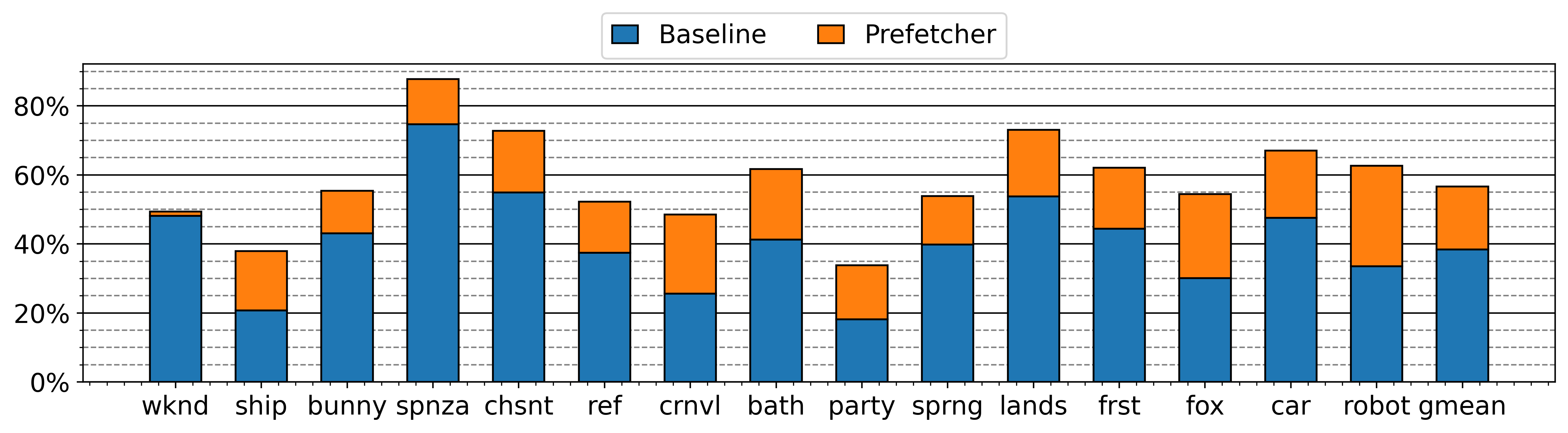}
            \label{avg_dram_bw}
            }\hfill
        \subfloat[Total number of DRAM reads and L2 writebacks with TTP normalized to baseline.]{
            \includegraphics[width=0.48\textwidth]{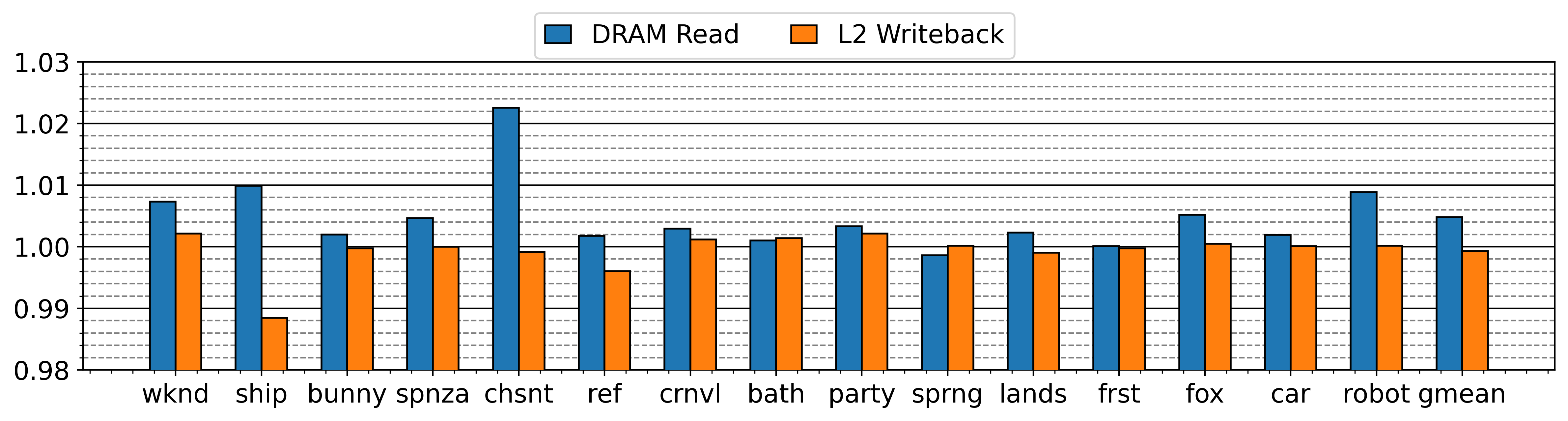}
            \label{dram_reads}
        }    
    % \begin{subfigure}[b]{0.48\textwidth}
    %     \centering
    %     \includegraphics[width=\textwidth]{figs/avg_dram_bw.png}
    %     \caption{Average DRAM bandwidth utilization.}
    %     \label{avg_dram_bw}
    %     \vspace{0.2cm}
    % \end{subfigure}
    % \hfill
    % \begin{subfigure}[b]{0.48\textwidth}
    %     \centering
    %     \includegraphics[width=\textwidth]{figs/dramreadwb.png}
    %     \caption{Total number of DRAM reads and L2 writebacks with TTP normalized to baseline.}
    %     \label{dramreadwb}
    % \end{subfigure}
    \caption{DRAM activity with and without (i.e., baseline) TTP.}
    \label{dramstats}
\end{figure}

To better understand the performance of ray tracing workloads, Figure \ref{rtcyclesdist} shows the averaged distribution of thread status while executing a \textit{trace\_ray} instruction (the methodology is in Section \ref{methodsection}).
The \textit{trace\_ray} instruction performs BVH traversal on the specialized GPU hardware, the RT unit. 
Within the RT unit, threads spend most of their time waiting for memory read requests to return, confirming that the memory wall is a significant bottleneck in ray tracing.
%From the figure, we can see that the ray tracing performance is dominated by the RT and MEM instructions. 
%, with most time spent waiting for small BVH node data from memory. 
To further analyze the memory performance of ray tracing workloads, we report the average DRAM bandwidth with and without our Tree Traversal Prefetcher (TTP) in Figure \ref{avg_dram_bw}. 
From the figure, we can see that most of the scenes do not fully utilize the DRAM bandwidth, suggesting ray tracing is mostly constrained by memory latency instead of bandwidth, also shown in previous work\cite{saed_rayn_2025}.
Additionally, underutilization of the bandwidth in the baseline suggests that there is sufficient headroom for a prefetcher.

In this paper, we aim to accelerate ray tracing by addressing its memory bottlenecks. 
We propose TTP to reduce the memory access latency. 
The key idea of TTP is to leverage the existing per-thread traversal stack in the RT unit for generating prefetches based on the tree traversal trend, which is predicted from the stack push/pop operation sequences. 
Our analysis shows that TTP generates highly accurate prefetches (i.e., the prefetched data are demanded in a timely manner) with minor hardware changes to the RT unit.  
As seen in Figure \ref{avg_dram_bw}, TTP introduces a bandwidth overhead of 18.22\% on average. 
On the other hand, the overall amount of data loaded from or written to DRAM remains nearly identical, as shown in Figure \ref{dram_reads}.
This shows that the increased bandwidth from TTP is due to the reduced execution time rather than extra data transfer from/to DRAM. 
Our results also show that TTP outperforms the state-of-the-art Treelet prefetcher\cite{treelet}, which was specifically designed for RT units. 

In summary, this paper makes the following contributions:
\begin{itemize}
    \item We propose TTP, a novel hardware prefetcher for ray tracing, which leverages the existing traversal hardware to prefetch data and hide the memory access latency. 
    \item We study the traversal trends across the 3D scenes and show that TTP is highly effective in various ray tracing workloads and outperforms the state-of-the-art Treelet prefetcher.
    \item Our simulation results show that TTP achieves up to 1.89x performance improvement with a geometric mean of 1.48x.
    \item TTP supports both DFS- and BFS-based traversal, offering a new performance optimization opportunity by selecting different traversal algorithms for different workloads.
\end{itemize}

\section{Background} \label{backgroundsec}
\subsection{Ray Tracing}
%Talk about ray tracing, bvh traversal, shader types
Ray tracing is a 3D rendering technique based on realistically modeling how light rays travel through space.
High-performance graphics application programming interfaces (APIs) such as Vulkan\cite{vulkanorg} and DirectX\cite{noauthor_directx_nodate} support ray tracing pipelines, which enable developers to build ray tracing applications.
Ray tracing pipelines introduce new programmable shader types such as ray generation(\textit{raygen}), closest-hit(\textit{chit}), any-hit(\textit{ahit}), intersection(\textit{rint)} and miss(\textit{miss}) shaders.
Raygen shader is the beginning of the pipeline, where \textit{primary} rays are generated by defining an origin and a direction in 3D space.
Rays generated from the user's position in the world toward the scene are called \textit{primary} rays. 
Rays that result from interactions with objects—such as reflections or refractions—are referred to as \textit{secondary} rays.
While the latest GPUs have built-in hardware support for testing intersections for triangles, procedural shapes such as spheres and cylinders require the \textit{intersection} shaders to check if the ray intersects the object.
If the intersected object is not opaque, then the \textit{any-hit} shader is called for every intersection.
Otherwise, at the end of the traversal, if the ray hits an object, the \textit{closest-hit} shader is called. 
If the ray misses all objects, the \textit{miss} shader is called. 

Two types of ray tracing are supported, \textit{closest-hit} and \textit{any-hit}, for various purposes and applications. % such as path tracing, ambient occlusion, or shadows.
Closest-hit rays search for the primitive that is closest to the ray origin, whereas any-hit rays terminate their search upon finding any primitive that the ray intersects.
Primary rays are always closest-hit rays; secondary rays are also closest-hit for path tracing, but any-hit for shadow and ambient occlusion shaders. 
Path tracing renders a 2D image of a 3D scene by tracing multiple rays per pixel and computing their resulting colors. 
Unlike hybrid approaches, which use ray tracing selectively for effects like shadows or ambient occlusion alongside rasterization, path tracing uses ray tracing exclusively to generate the entire frame.
Shadow shaders determine the amount of direct light that reaches objects in a scene by tracing rays from the objects to the light sources.
Ambient occlusion shaders estimate the amount of ambient light that reaches objects by tracing rays in random directions from the objects and calculating the degree of occlusion.
Among the use cases of ray tracing, path tracing is the most intensive and time consuming workload.
Typically, 4 to 16 bouncing rays are traced per pixel to determine its final color, compared to 2 or 4 rays traced for shadows and ambient occlusion. 
In addition, path tracing rays are always closest-hit, whereas other workloads use any-hit rays which are faster to trace.

\subsection{BVH Traversal}
To accelerate ray tracing, a tree-like structure called Bounding Volume Hierarchy (BVH) is employed, which consists of Axis-Aligned Bounding Boxes(AABB) that hierarchically capture scene primitives \cite{bvh_1}. 
Rays first test outer boxes before deeper hierarchy levels, reducing the number of intersection tests.
Primitives are leaf nodes of the tree, which are at the bottom.
As rays test intersections and traverse deeper into the tree, they reach the leaf nodes and test against the primitives to determine if there is a hit.
If a hit is found, remaining nodes are skipped if they are farther than the closest-hit primitive, which saves time.

As with any tree search problem, the tree can be traversed in depth-first or breadth-first fashion.
For ray tracing, depth-first search (DFS) is often preferred as it quickly descends deeper and identifies a hit primitive, which allows the remaining nodes to be skipped if they are farther.
A stack (last-in first-out or LIFO) data structure stores node addresses (not the node data itself) that will be read from memory next.
Breadth-first search (BFS) is also a viable option. However, BFS processes nodes on the same level before going deeper, and takes longer to identify the first primitive hit. 
Therefore, it loses the opportunity to skip farther nodes.
BFS uses a queue (first-in first-out or FIFO) structure to keep track of nodes, unlike the stack in DFS.

\begin{figure}[]
\centerline{\includegraphics[width=0.4\textwidth]{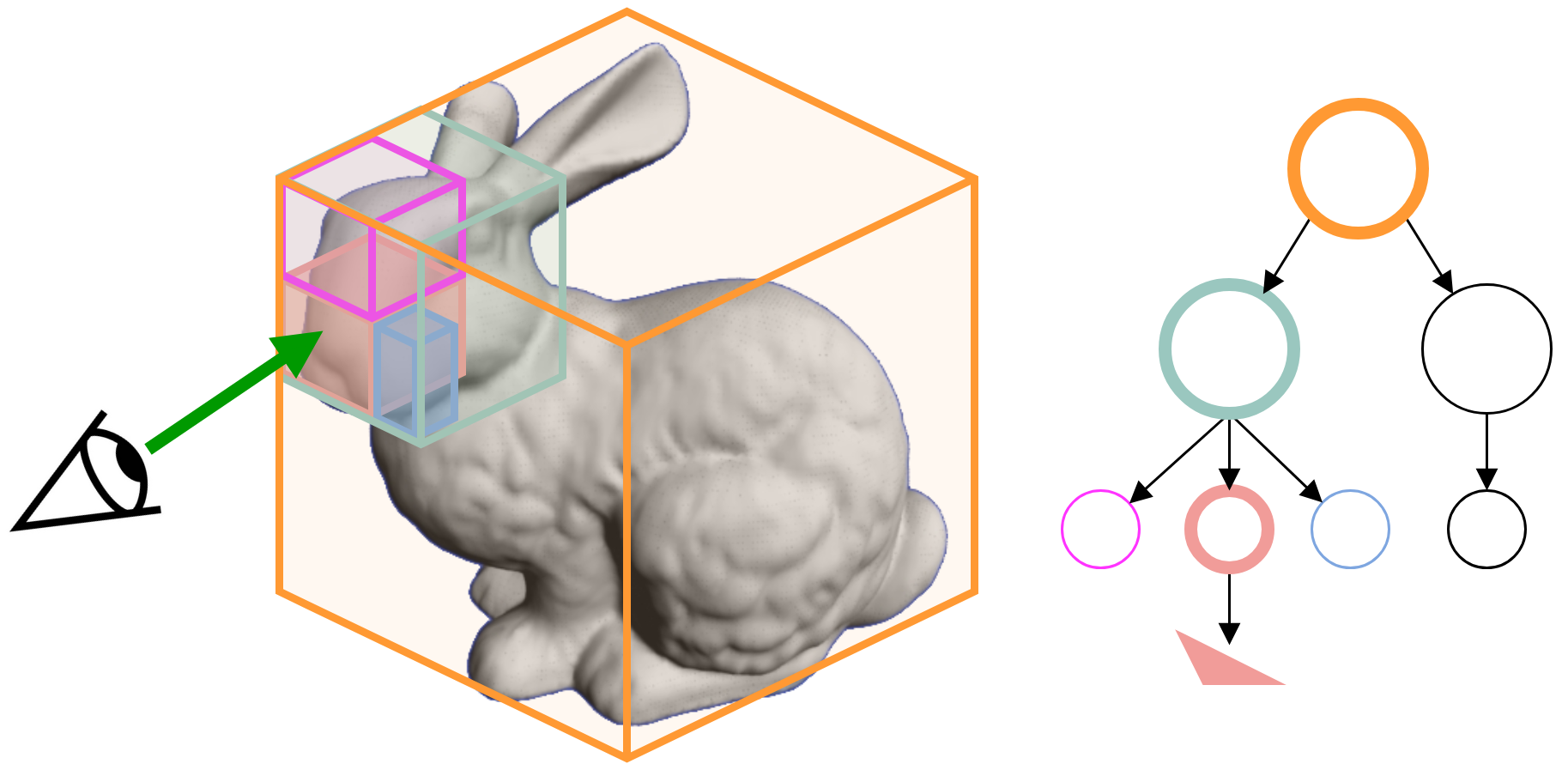}}
\caption{Simplified BVH structure for Stanford Bunny. The green arrow represents the ray coming out of the camera, i.e. the eye. The ray intersects with the triangle inside the pink box.}
\label{bunny_bvh}
\end{figure}

Figure \ref{bunny_bvh} visualizes a simplified BVH structure and the corresponding tree for the Bunny object\cite{noauthor_stanford_nodate}.

\subsection{GPU Architecture and Hardware Accelerated Ray Tracing}
\begin{figure*}[]
\centerline{\includegraphics[width=0.91\textwidth]{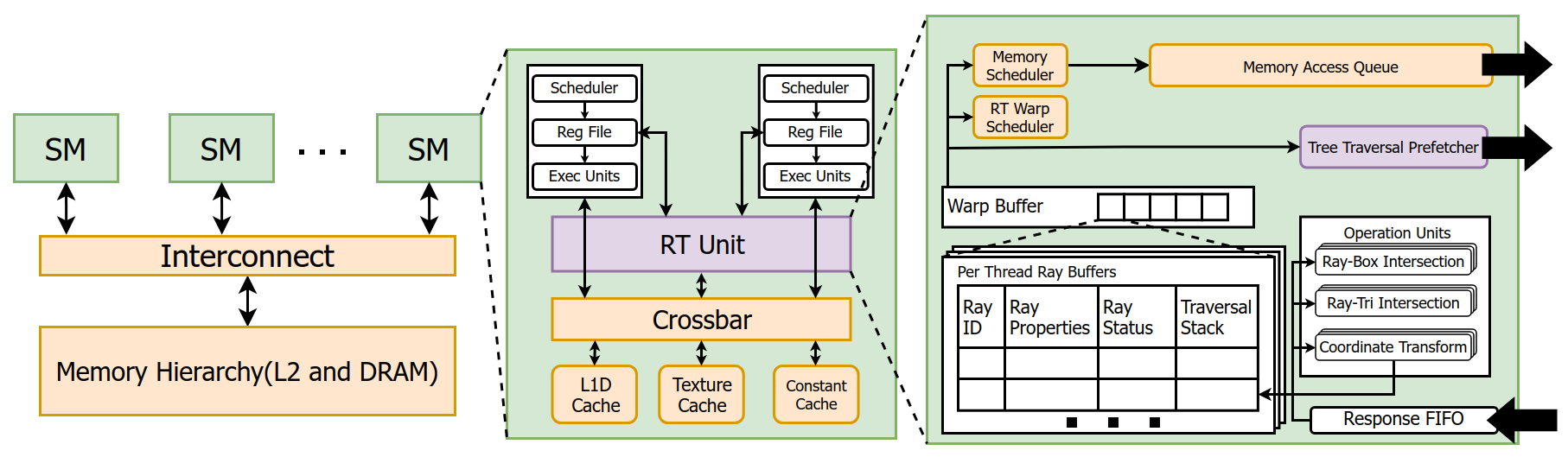}}
\caption{Diagram of the GPU model used in this study. Purple blocks indicate the modified components. Redrawn from \cite{vksim}.}
\label{gpu_diagram}
\end{figure*}

GPUs are Single Instruction Multiple Thread (SIMT) computers.
They support thousands of hardware threads that run in parallel, thus yielding massive computational throughput\cite{accelsim}.
Figure \ref{gpu_diagram} shows the baseline GPU model \cite{vksim}, with our proposed modifications highlighted in purple.
A GPU consists of an array of Streaming Multiprocessors (SMs), as termed by Nvidia.
SMs act as the main compute units where shader programs are executed. 
Each SM houses a register file, schedulers, hundreds of execution lanes, and an L1D cache.
Shader programs are executed by groups of 32 threads, called "warps" by Nvidia, which operate in a lock-step manner. 
All the SMs are connected to an L2 cache via an interconnect, which is in turn connected to the DRAM memory controllers.

The latest GPUs also incorporate an RT unit in the SMs\cite{ada_arch}\cite{ampere_arch}.
RT units are introduced to accelerate the BVH tree traversal process, which accounts for the majority of ray tracing execution time \cite{vksim}.
Tree traversal involves fetching the BVH tree nodes from memory, testing for intersections, and fetching subsequent nodes based on the intersection results.
The RT unit model used in this study is shown in Figure \ref{gpu_diagram}.
A warp buffer stores the per thread traversal stack and ray metadata.
At each cycle, a warp from the warp buffer is selected and a memory request from that warp is served.
Duplicate requests within the same warp are coalesced into one request, and inserted into the memory access queue, which sends them to the memory hierarchy (i.e., L1 cache).
In the meantime, memory responses are inserted into the response FIFO. 
Data in the response FIFO are fed into math units which perform ray-box, ray-triangle intersection tests or coordinate transformations depending on the node type. 
Finally, the warp buffers are updated and new node addresses are pushed to stacks based on intersection test results.
Traversal continues until all threads in the warp have emptied their stacks (closest-hit), or found a hit (any-hit).

RT units accelerate BVH traversal by streamlining the memory reads and intersection tests.
BVH traversal is carried out within the RT unit by a single CISC-like instruction, i.e., the \textit{trace\_ray} instruction. 
Dedicated memory schedulers and fixed-function intersection units inside the RT unit handle memory reads and primitive intersection tests internally, eliminating the need for explicit memory and compute instructions in the shader code.

During BVH traversal, GPU threads spend most of their time waiting for small pieces of tree data to arrive from memory \cite{treelet}.
As studied in the previous work \cite{treelet}, prefetchers can significantly mitigate this bottleneck by predicting which nodes are going to be read using heuristics and prefetching those nodes just before the threads need them.
%In this work, we propose a novel prefetcher for tree traversal that works by monitoring the per-thread traversal stack readily available in the RT unit. As tree nodes are "popped", i.e. removed, from the stack, we prefetch the subsequent nodes in the traversal stack.

\subsection{Hardware Prefetchers}
Hardware prefetchers, primarily studied for CPUs, include simple types like stream and stride prefetchers, which load data from consecutive addresses or detect strided memory accesses typically occurring in loops to reduce capacity and compulsory cache misses \cite{fu_stride_1992}\cite{jouppi_improving_1990}. % but fail with irregular accesses. 
Global History Buffer (GHB) prefetchers maintain a FIFO table of recent cache misses and use a linked list to traverse addresses for prefetching\cite{nesbit_data_2004}. 
Indirect Memory Prefetchers (IMP) predict indirect memory accesses using address offsets and detect patterns with an Indirect Pattern Detector (IPD)\cite{yu_imp_2015}. 
While GHB and IMP handle irregular patterns, they struggle with the randomness of ray tracing workloads. Graph prefetchers aim to accelerate graph search algorithms but are less effective for the irregular nature of ray tracing\cite{ainsworth_graph_2016}.
Recent work \cite{treelet} introduced the Treelet Prefetcher specifically designed for GPU ray tracing, which builds and prefetches treelets within the BVH tree. It requires a custom traversal algorithm and a rearranged BVH tree for optimal performance.

\section{Traversal Trend in Ray Tracing} \label{section3}
We first study the BVH traversal trend in our ray tracing workloads to deeply understand the underlying traversal behavior and uncover prefetching opportunities. 
In particular, we look at the traversal stack activity to identify patterns which can be exploited.

As mentioned before, there are two types of hits that can be searched: closest-hit and any-hit.
For closest-hit, the traversal is typically longer, as the whole tree must be covered to correctly identify the closest-hit primitive. 
This can lead to repetitive downward and upward traversal through the BVH tree, depending on ray origin and tree structure.
Any-hit, on the other hand, requires identifying an intersected primitive regardless of its distance.
We focus on \textit{closest-hit} rays, as they are used in path tracing, which is the most intensive ray tracing workload.
Algorithm \ref{trav_alg} shows how DFS is carried out to find the closest-hit primitive.
First, the root node is pushed to stack if its AABB is intersected (lines 1-2).
Then, as long as the stack is not empty, node address at the top is popped and read from memory (lines 4-5).
If it is an internal node, it contains the addresses and AABB coordinates of its child nodes, which are tested for intersection and pushed to stack if hit (lines 6-10).
Otherwise, it is a leaf node which can be a triangle or a procedural geometry necessitating an intersection shader to be executed.
If an intersected primitive is found, the closest-hit distance(\textit{min\_thit}) is updated (line 15).
At the end of traversal, if a hit is found, registers are updated to indicate the hit, and geometry metadata is stored in memory to be read by the closest-hit or any-hit shaders.

\begin{algorithm}
\small
\caption{BVH Traversal using DFS to find the closest-hit primitive}
\label{trav_alg}
\KwIn{$ray, root\_node\_addr$}
\KwOut{$closest\_node$}
\If{$ray$ intersects $root\_node$}{
    $stack.push(root\_node\_addr)$\;
}
\While{$stack\ is\ not\ empty$}{
    $node\_addr \gets stack.pop()$\;
    $node \gets mem\_read(node\_addr)$\;
    \eIf{$node\ is\ internal\ node$}{
        \For(\tcp*[h]{6-ary tree}){$i = 0$ \textbf{to} $5$} {
            $thit \gets ray\_box\_test(ray,node.child[i].AABB)$\;
            \If{$thit<min\_thit$}{
                $stack.push(node.child[i].addr)$\;
            }
        }
    }(\tcp*[h]{leaf node}){
        $thit \gets ray\_triangle\_test(ray,node)$\;
        \If{$thit<min\_thit$} {
            $closest\_node \gets node$\;
            $min\_thit \gets thit$\;
        }
    }
}
$return\ closest\_node$\;
\end{algorithm}

\subsection{Prefetch Opportunities in DFS}
In DFS, nodes are pushed to and popped from top of stack.
Consequent pushes indicate downward traversal towards the leaf nodes, while pops indicate upward traversal towards the root or across the same level. 
When consecutive pushes happen (downward traversal), it is difficult to predict which nodes will be needed, making prefetching highly challenging and speculative in nature, leading to wasted bandwidth. 
Since node pushing is determined by the results of intersection tests, it is often too late to prefetch the nodes by the time they are needed.
In other words, top of stack may change depending on the result of the current intersection test, so the next node to be read cannot be known until the test has completed.
On the contrary, when consecutive pops happen (upward or same-level traversal), addresses of nodes to be read next are readily available in the traversal stack, and they can be trivially retrieved in a timely manner. This is the key insight driving our prefetcher design. 

\begin{figure}[]
    \subfloat[BVH Tree traversal with depth-first search. The ray eventually hits nodes $O$ and $E$. Green arrows indicate a pop from the traversal stack followed by a push of its children. For example, node $A$ is popped and read from memory. $A$'s node data contains AABB coordinates of its children $B$, $C$, and $D$, among which $B$ and $D$ test positive for intersection, and therefore pushed to stack. Red arrows indicate a pop from the stack followed by another pop. For example, $O$ is popped after $P$ is popped.]{
        \centering
        \includegraphics[width=0.45\textwidth]{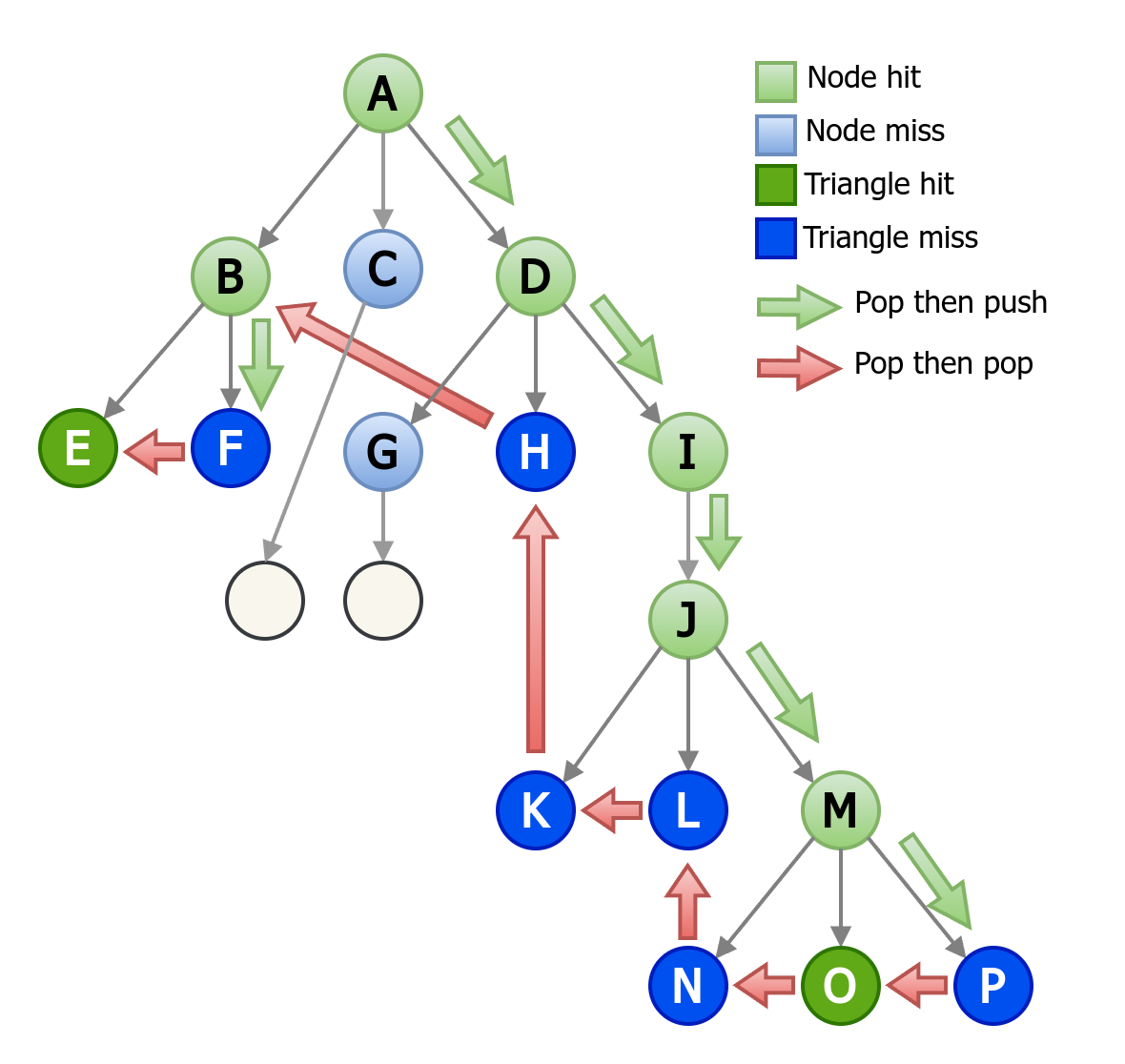}
        \label{bvh_tree}
    }\hfill
    \vspace{-0.6cm}
    \subfloat[Traversal stack right after $O$ is popped. The stack contains the nodes' addresses. $N$ is used instead of \&$N$ for simplicity.]{
        \centering
        \hspace{0.14\textwidth}% Add left spacing
        \includegraphics[width=0.16\textwidth]{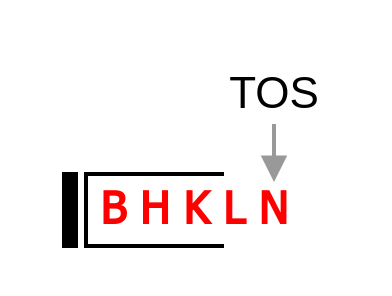}
        \hspace{0.14\textwidth}% Add right spacing  
        \label{stack}
    }
    \caption{Example DFS BVH traversal and the corresponding traversal stack.}
    \label{tree_traversal}
\end{figure}

Figure \ref{bvh_tree} shows an example BVH tree traversal using the DFS algorithm. 
The traversal starts with the intersection test at the root node $A$. 
We assume that the ray eventually hits leaf nodes $O$ and $E$. 
Therefore, when node $A$ is popped and read from memory, the intersection test would show the ray hits the bounding boxes of $B$ and $D$, whose addresses are pushed to stack.
Then the address of $D$ is popped and read from memory. 
Its children are tested for intersection and $H$, $I$ are pushed to stack.
This traversal continues until the whole tree is traversed, and nodes $O$ and $E$ are found as triangles that are hit. 
Note that the children of $G$ and $C$ are ignored because their bounding boxes are not hit and they are not pushed to the stack. 
Addresses of nodes $P$, $N$, $L$, $K$, $H$, and $F$ are pushed to the stack because the ray intersects their bounding boxes, but misses the triangle inside them.
Figure \ref{stack} shows the content of the thread’s traversal stack right after the address of $O$ is popped.

From this example, we make an important observation on the traversal trend in DFS, which is the frequent down-and-up traversal through the tree. 
In other words, a series of pop-push operations (traversing down the tree) are typically followed by a series of pop operations (traversing up the tree). 
In the example shown in Figure \ref{bvh_tree}, a tree branch, $A\rightarrow D\rightarrow I\rightarrow J\rightarrow M\rightarrow P\rightarrow O\rightarrow N$, is traversed until its tip is reached, then the upward traversal, $N\rightarrow L\rightarrow K\rightarrow H\rightarrow B$, begins.
As the traversal stack already contains the addresses of these nodes, as shown in Figure \ref{stack}, we can prefetch them when we observe consecutive pops, meaning the thread is traversing the tree upward.

Note that although the node addresses (e.g., $N,L,K,H,B$) are in the stack, these nodes themselves have not been fetched and processed yet by the current thread (or ray). 
Therefore, prefetching them can reduce both cold misses if no other prior rays (or threads) have fetched them, and capacity or conflict misses if they were fetched by other threads but evicted from the caches later on.

\begin{figure}[]
    \centering
        \subfloat[Pop streak frequencies.]{
        \centering
        \includegraphics[width=0.48\textwidth]{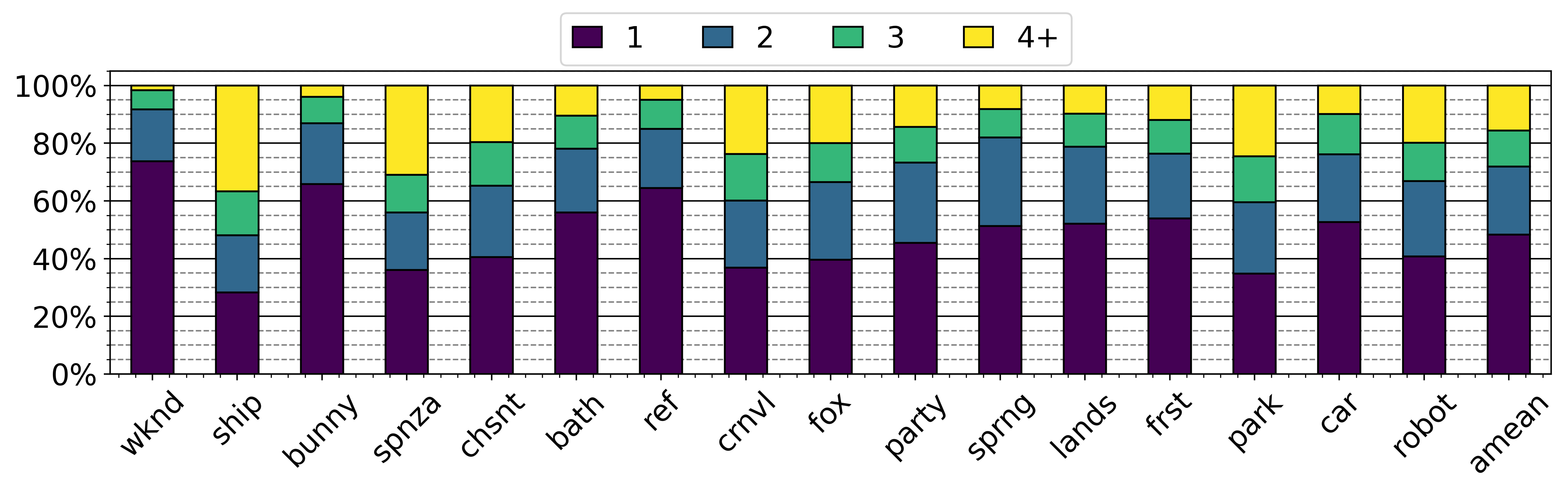}
        \label{pop_freqs}
        \vspace{0.2cm}
    }
    \hfill
    \subfloat[Composition of RT read misses in terms of pop streaks. The misses are those generating DRAM traffic, i.e., missing in both L1 and L2.]{
        \centering
        \includegraphics[width=0.48\textwidth]{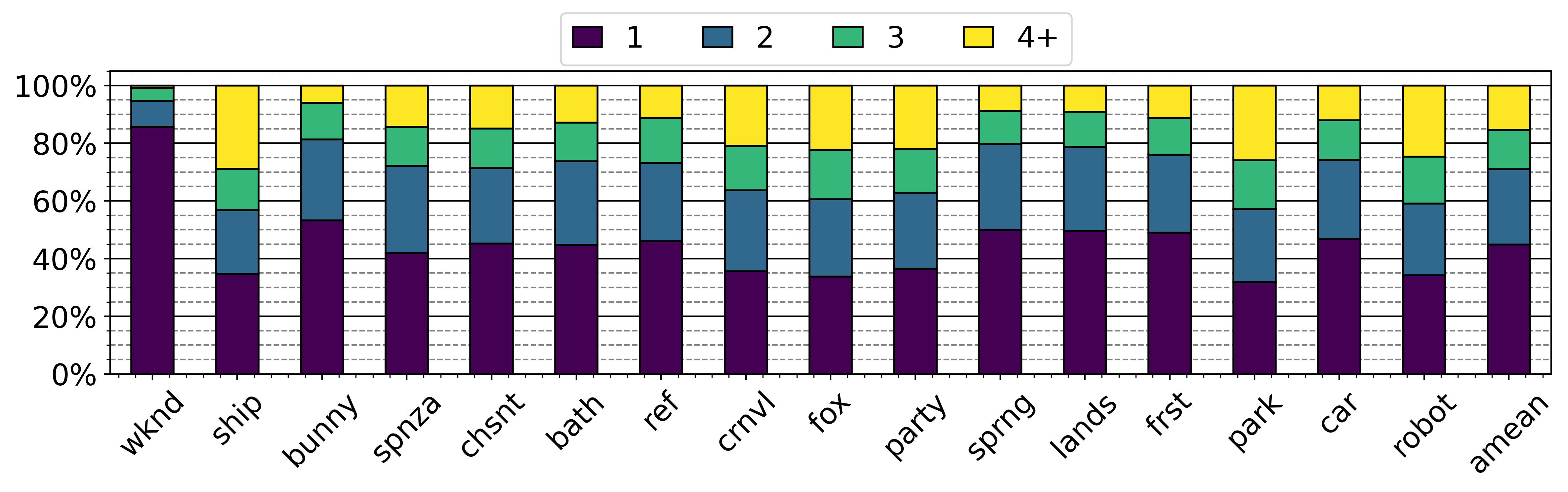}
    }
    \caption{Analysis of pop streaks.}
    \label{pop_miss_rates}
\end{figure}

To see the impact of such traversal stack pop streaks, we extract traversal stack activity from the simulator.
Note that during BVH traversal, every memory read request is a \textit{pop} from a traversal stack.
Therefore, we categorize the cache misses from RT reads by the position of the pop within a streak—i.e., whether it is the 1st, 2nd, 3rd, or 4th+ consecutive pop. 
The 1st pop (i.e., no consecutive pops) means downward traversal while consecutive pops(2, 3, 4+) correspond to upward or same-level traversal.
We quantify both the frequency of these pop streaks and their cache miss rates in Figure \ref{pop_miss_rates}.
Although single pops, i.e. a pop followed immediately by a push, make up most of the stack activity, longer pop streaks make up a significant portion of all pops. 
Consequently, and more importantly, these longer pop streaks account for a large portion of all cache misses, making them prime candidates for prefetching.

\subsection{Prefetch Opportunities in BFS}
BFS is another option for tree traversal, where nodes at the same level are processed before descending deeper into
the tree. 
In this case, the traversal stack would operate as a FIFO queue rather than a LIFO stack. Taking the BVH tree in Figure \ref{tree_traversal} as an example, after node A is popped and accessed, both nodes B and D will be pushed into a queue (i.e., queue content: B,D). 
After B is popped and accessed, its children E and F will be pushed into the queue after node D (i.e., queue content: D, E, F). Next, node D will be popped and accessed as it is the head of the queue.

While BFS is commonly used in traversing graphs \cite{buluc_parallel_2011}\cite{luo_effective_2010}\cite{merrill_scalable_2012}, it is less effective for ray tracing compared to DFS, as it takes longer to identify the first hit and cannot skip as many nodes. 
Table \ref{avg_nodes_table} shows the average number of nodes per ray visited by BFS and DFS for path tracing shaders, which trace closest-hit rays.

However, BFS has one important advantage over DFS: predictability.
Because BFS uses a FIFO queue, the next node that will be read is known as long as the queue is not empty.
New nodes are pushed to the tail of the queue, and the next node is popped from the head.
This observation makes BFS an appealing candidate for prefetching as studied in previous work\cite{ainsworth_graph_2016}.
The key to accurate prefetching with BFS is timing.
Although the address of the next node is readily available regardless of upward or downward traversal, prefetch must be done at the right time to avoid being too early or late. 

With BFS, the traversal trend is always traversing the nodes on the same level before moving on to the next level. 
A prefetch opportunity exists as long as the queue is not empty. 
Following the example in Figure \ref{bvh_tree}, when the node B is to be popped, the queue content is the addresses of node B and D. 
As a result, when B is sent to the memory access queue, D can be prefetched. 
If both are cache misses, their latency can be overlapped. 

\begin{figure}[]
\centerline{\includegraphics[width=0.48\textwidth]{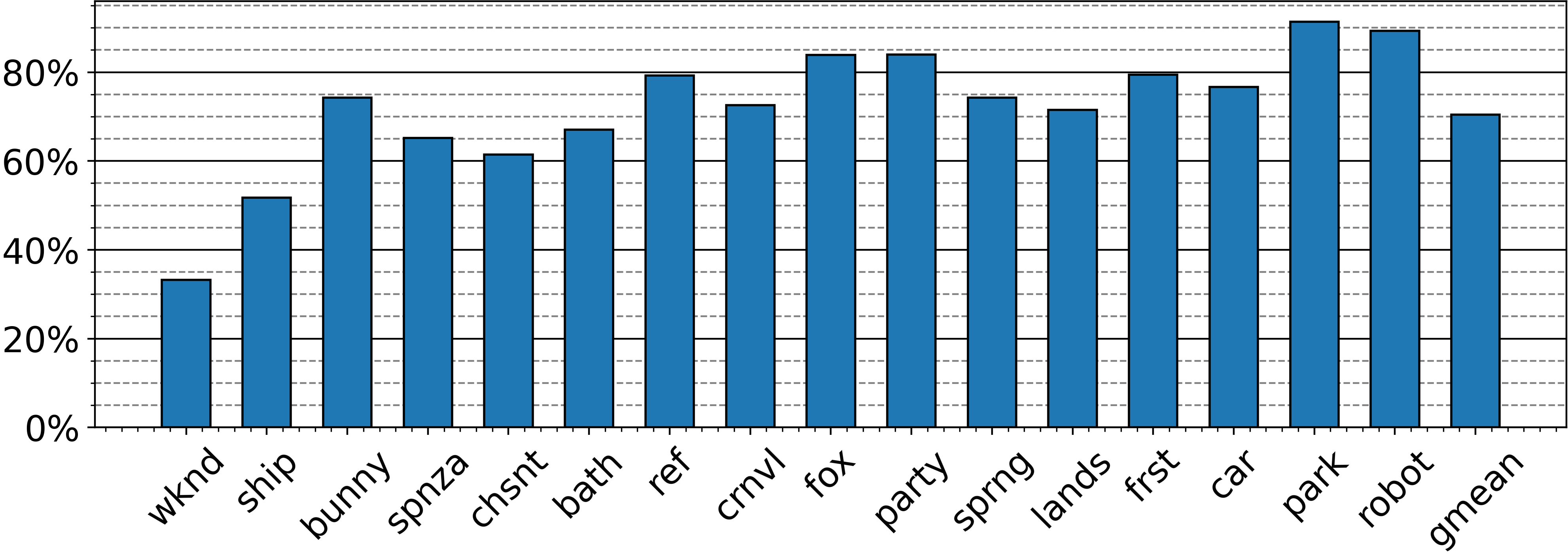}}
\caption{Percentage of RT read misses where the node was in the traversal queue when a previously read node missed in caches.}
\label{bfs_trend}
\end{figure}

In Figure \ref{bfs_trend}, we show the percentage of RT read misses that satisfy the conditions, i.e., the queue is not empty and the demand access is a cache miss.
In theory, these results show that, on average, 70.42\% of all queue pops can be followed with an accurate prefetch, i.e. an upperbound on the prefetcher's effectiveness.
In practice, scheduling uncertainties and cache sizes will limit the effectiveness of the prefetcher.

\begin{table}[] 
\centering
\caption{Comparison of DFS and BFS traversal in terms of average and maximum nodes visited per ray. Positive Diff means BFS visits more nodes.}
\begin{tabular}{|l|lll|lll|} 
\hline
       & \multicolumn{3}{c|}{\textbf{Average Nodes Per Ray}}                                      & \multicolumn{3}{c|}{\textbf{Max Nodes Per Ray}}                                          \\ \hline
Scenes & \multicolumn{1}{c|}{DFS} & \multicolumn{1}{c|}{BFS} & \multicolumn{1}{c|}{\textbf{Diff}} & \multicolumn{1}{c|}{DFS} & \multicolumn{1}{c|}{BFS} & \multicolumn{1}{c|}{\textbf{Diff}} \\ \hline
WKND   & \multicolumn{1}{c|}{13.3}  & \multicolumn{1}{c|}{13.3}  & \multicolumn{1}{c|}{0.0\%}       & \multicolumn{1}{c|}{73}     & \multicolumn{1}{c|}{73}     & \multicolumn{1}{c|}{0.0\%}        \\ \hline
SHIP   & \multicolumn{1}{c|}{43.3}  & \multicolumn{1}{c|}{46.6}  & \multicolumn{1}{c|}{7.6\%}       & \multicolumn{1}{c|}{240}    & \multicolumn{1}{c|}{245}    & \multicolumn{1}{c|}{2.1\%}        \\ \hline
BUNNY  & \multicolumn{1}{c|}{11.3}  & \multicolumn{1}{c|}{12.5}  & \multicolumn{1}{c|}{10.6\%}      & \multicolumn{1}{c|}{141}    & \multicolumn{1}{c|}{175}    & \multicolumn{1}{c|}{24.1\%}       \\ \hline
SPNZA  & \multicolumn{1}{c|}{38.0}  & \multicolumn{1}{c|}{51.5}  & \multicolumn{1}{c|}{35.5\%}      & \multicolumn{1}{c|}{335}    & \multicolumn{1}{c|}{385}    & \multicolumn{1}{c|}{14.9\%}       \\ \hline
CHSNT  & \multicolumn{1}{c|}{59.0}  & \multicolumn{1}{c|}{137.4}  & \multicolumn{1}{c|}{132.9\%}      & \multicolumn{1}{c|}{274}    & \multicolumn{1}{c|}{574}    & \multicolumn{1}{c|}{109.5\%}       \\ \hline
BATH   & \multicolumn{1}{c|}{18.4}  & \multicolumn{1}{c|}{21.1}  & \multicolumn{1}{c|}{14.7\%}      & \multicolumn{1}{c|}{249}    & \multicolumn{1}{c|}{278}    & \multicolumn{1}{c|}{11.6\%}       \\ \hline
REF    & \multicolumn{1}{c|}{11.3}  & \multicolumn{1}{c|}{12.7}  & \multicolumn{1}{c|}{12.4\%}      & \multicolumn{1}{c|}{235}    & \multicolumn{1}{c|}{291}    & \multicolumn{1}{c|}{23.8\%}       \\ \hline
CRNVL  & \multicolumn{1}{c|}{43.8}  & \multicolumn{1}{c|}{51.4}  & \multicolumn{1}{c|}{17.4\%}      & \multicolumn{1}{c|}{487}    & \multicolumn{1}{c|}{492}    & \multicolumn{1}{c|}{1.0\%}        \\ \hline
FOX    & \multicolumn{1}{c|}{84.0}  & \multicolumn{1}{c|}{117.2} & \multicolumn{1}{c|}{39.5\%}      & \multicolumn{1}{c|}{568}    & \multicolumn{1}{c|}{874}    & \multicolumn{1}{c|}{53.7\%}       \\ \hline
PARTY  & \multicolumn{1}{c|}{34.8}  & \multicolumn{1}{c|}{37.4}  & \multicolumn{1}{c|}{7.5\%}       & \multicolumn{1}{c|}{551}    & \multicolumn{1}{c|}{551}    & \multicolumn{1}{c|}{0.0\%}        \\ \hline
SPRNG  & \multicolumn{1}{c|}{31.4}  & \multicolumn{1}{c|}{38.7}  & \multicolumn{1}{c|}{23.2\%}      & \multicolumn{1}{c|}{204}    & \multicolumn{1}{c|}{291}    & \multicolumn{1}{c|}{42.6\%}       \\ \hline
LANDS  & \multicolumn{1}{c|}{30.2}  & \multicolumn{1}{c|}{38.9}  & \multicolumn{1}{c|}{28.8\%}      & \multicolumn{1}{c|}{527}    & \multicolumn{1}{c|}{535}    & \multicolumn{1}{c|}{1.5\%}        \\ \hline
FRST   & \multicolumn{1}{c|}{35.2}  & \multicolumn{1}{c|}{49.7}  & \multicolumn{1}{c|}{41.2\%}      & \multicolumn{1}{c|}{355}    & \multicolumn{1}{c|}{943}    & \multicolumn{1}{c|}{165.6\%}      \\ \hline
CAR    & \multicolumn{1}{c|}{41.1}  & \multicolumn{1}{c|}{53.3}  & \multicolumn{1}{c|}{29.7\%}      & \multicolumn{1}{c|}{303}    & \multicolumn{1}{c|}{582}    & \multicolumn{1}{c|}{92.4\%}       \\ \hline
PARK   & \multicolumn{1}{c|}{185.5} & \multicolumn{1}{c|}{272.3} & \multicolumn{1}{c|}{46.8\%}      & \multicolumn{1}{c|}{1753}   & \multicolumn{1}{c|}{2071}   & \multicolumn{1}{c|}{18.1\%}       \\ \hline
ROBOT  & \multicolumn{1}{c|}{102.5} & \multicolumn{1}{c|}{166.6} & \multicolumn{1}{c|}{62.6\%}      & \multicolumn{1}{c|}{1516}   & \multicolumn{1}{c|}{1841}   & \multicolumn{1}{c|}{21.4\%}       \\ \hline
\textbf{AVERG}  & \multicolumn{1}{c|}{49.0} & \multicolumn{1}{c|}{70.0} & \multicolumn{1}{c|}{42.9\%}      & \multicolumn{1}{c|}{487.8}   & \multicolumn{1}{c|}{636.9}   & \multicolumn{1}{c|}{30.6\%}       \\ \hline
\end{tabular}
\label{avg_nodes_table}
\end{table}

\section{Tree Traversal Prefetcher} \label{prefsection}

\subsection{Generating Prefetches with DFS}
\begin{figure}[]
\centerline{\includegraphics[width=0.44\textwidth]{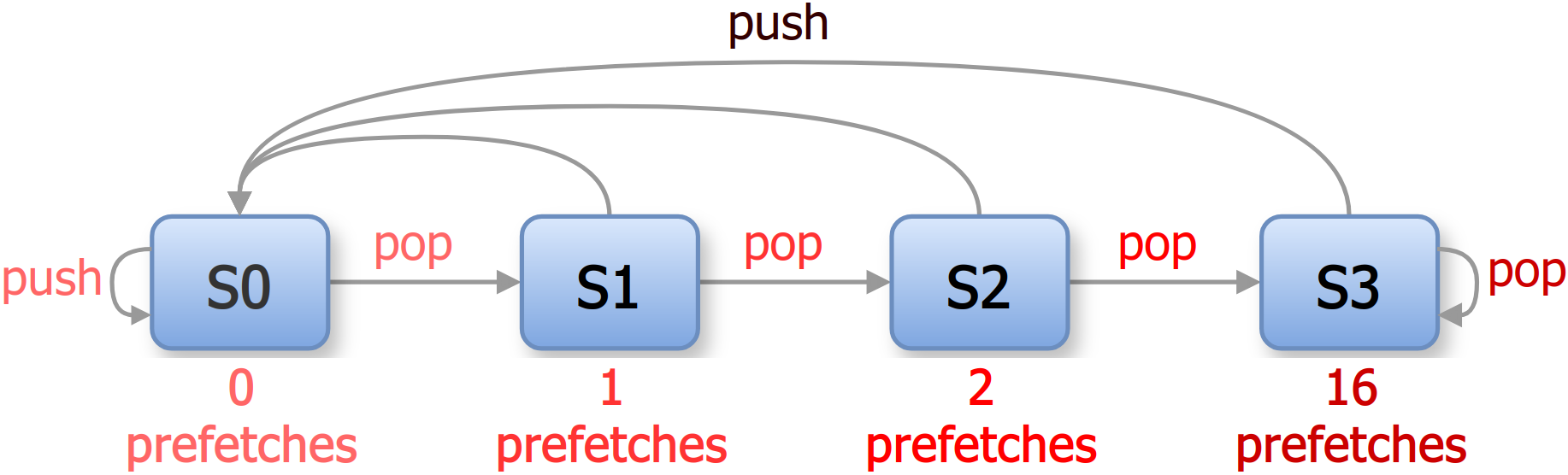}}
\caption{State machine that generates prefetches.}
\label{state_machine}
\end{figure}

For DFS-based ray tracing, we propose to generate prefetches when the traversal trend is upward along the tree. To do so, we monitor the actions of the traversal stack: pushes and pops.
With the first \textit{pop} after a \textit{push}, we generate 1 prefetch, which is the node at the top of the stack.
With the second consecutive \textit{pop}, we generate 2 prefetches using the top two nodes in the stack, and with the third \textit{pop}, we are more confident about the trend and can be more aggressive in generating prefetches, e.g., by issuing up to 16 prefetches using the top 16 nodes in the stack.
A \textit{push} would reset the \textit{pop} streak and stop generating prefetches.
Figure \ref{state_machine} shows the state machine that implements this operation. This state machine can be implemented using 2 bits. There is no need to predict prefetch addresses as we simply use the content of the traversal stack. As each thread has its own traversal stack, our proposed TTP is a per-thread prefetching engine, sending prefetches when the warp is selected by the scheduler. 

In the example shown in Figure \ref{bvh_tree}, $O$ would be prefetched when $P$ is popped from the stack (state S1), then $N$ and $L$ would be prefetched after $O$ is popped (state S2).

The state machine can be adjusted for more flexibility, including the number of prefetches in each state and the number of states. 
However, as this state machine is one per traversal stack and, therefore, per thread, increasing the number of states may increase the hardware cost. Therefore, this paper uses the design in Figure \ref{state_machine} and studies the sensitivity of the parameters in our experiments.

We choose not to prefetch when a thread traverses down the tree. 
The reason is that we need to predict both intersection test results at different nodes and the descendent node addresses. 
Unless the prediction accuracy is high, such a scheme may incur high bandwidth/energy overhead and/or slowdown due to mispredicted paths. 
In contrast, prefetching for upward traversal does not require any address predictions. 

\subsection{Generating Prefetches with BFS}
We propose a simpler prefetching scheme for BFS.
With every pop from the traversal queue, up to $N$ nodes are prefetched starting from the head.
The parameter \textit{N} determines the prefetch distance — that is, how far ahead in the queue the nodes are fetched.
A larger \textit{N} results in prefetching nodes that will be accessed further in the future.
We explore with different values for \textit{N} and report the results in Section \ref{bfsressec}.

\subsection{Implementation}
\begin{figure}[]
\centerline{\includegraphics[width=0.3\textwidth]{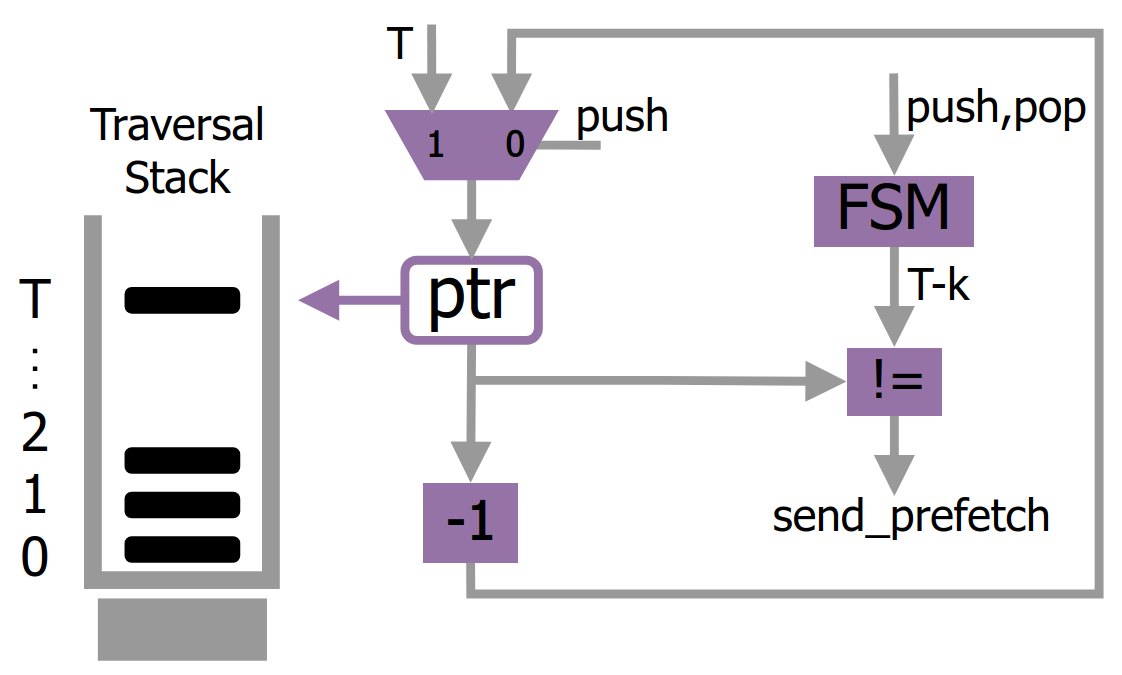}}
\caption{TTP implementation. Top of the stack is denoted by $T$, and the bottom is $0$. Purple blocks indicate newly added hardware structures. Only per-thread structures are shown.}
\label{ttp_impl}
\end{figure}

Figure \ref{ttp_impl} shows the implementation of TTP.
Due to the simplicity of the underlying algorithm, very little hardware is required.
We add an additional (2-bit) field to the warp buffer to implement the finite-state machine of each thread.
Push and pop actions in the traversal stack orchestrates the finite-state machine, which in turn calculates the $T-k$ value, where $T$ denotes the top of stack, and $k$ is the prefetch distance, i.e. $1$, $2$ or $16$.
We add a pointer that points to the next address in the stack that will be prefetched.
If a push happens, it is reset to $T$.
If a pop happens, FSM is updated, which updates the $T-k$ value.
Otherwise, with every prefetch, the pointer is decremented by one, which moves it to the next address in the stack.
When the pointer reaches $T-k$, as determined by the comparator, prefetching stops. The pointer only resets to \textit{T} upon a stack push, which prevents repeating prefetches from consecutive pops. Alternatively, each stack entry can have a flag bit to record whether the node address has been prefetched to avoid same entries being prefetched multiple times.% until the pointer is reset to $T$ with a push, or $T-k$ moves due to a pop.

By default, prefetch requests are sent when there are no demand reads, i.e. demand reads take priority. 
In Section \ref{agressec}, we report results with different arbitration schemes where prefetches take priority if no prefetches were sent recently.
Since the Vulkan-sim GPU model uses a sector cache model with sector size of 32B, prefetch requests for nodes that are larger than 32B are broken down into 32B chunks and sent one at a cycle, similar to demand reads.

When TTP is used with BFS, the prefetch distance $k$ is a fixed value, $N$, rather than determined by the finite-state machine. Our experiments in Section \ref{bfsressec} analyze the choice of this parameter.

\section{Methodology} \label{methodsection}
We extend Vulkan-sim 2.0\cite{vksim} to model our proposed TTP, and host the open-source code at github.com/yavuz650/vulkan-sim.
We use GPUWattch\cite{gpuwattch} for power analysis, which is included in the Vulkan-sim codebase.

The primary metric that we use for performance evaluation is the ratio of total number of simulation cycles, i.e., $cycles\_baseline/cycles\_prefetcher$.
Baseline is the default RT unit that is shipped with Vulkan-sim.

Lumibench has 16 scenes with increasing geometric complexity, as summarized in Table \ref{scene_table}.
15 out of 16 scenes finish simulation without errors at 128x128 resolution. The
\textbf{park} scene times out after 72 hours.
Due to this, simulations for the \textbf{park} scene are ran at 64x64 resolution instead, which run into completion.

\begin{table*}[]
\caption{Benchmark scenes from LumiBench\cite{lumibench}. %\textbf{park}, \textbf{car} and \textbf{robot} are too big for our simulations and would not finish. 
Scene stats taken from \cite{treelet}.}
\begin{tabular}{|l|c|c|c|c|c|c|c|c|}
\hline
Scene  &   \includegraphics[width=0.085\textwidth]{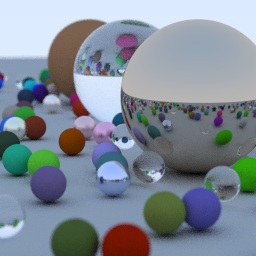} & \includegraphics[width=0.085\textwidth]{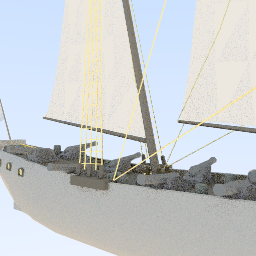} & \includegraphics[width=0.085\textwidth]{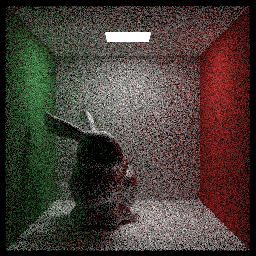} & \includegraphics[width=0.085\textwidth]{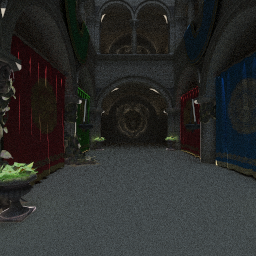} & \includegraphics[width=0.085\textwidth]{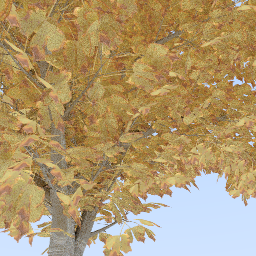} & \includegraphics[width=0.085\textwidth]{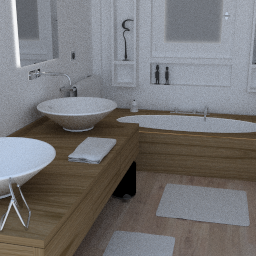} & \includegraphics[width=0.085\textwidth]{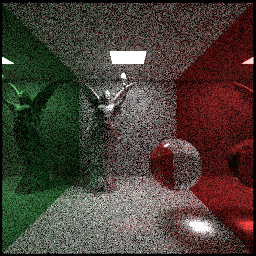} & \includegraphics[width=0.085\textwidth]{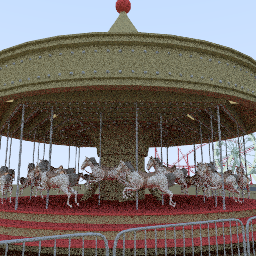}  \\ \hline
Label     & \textbf{\normalsize wknd} & \textbf{\normalsize ship} & \textbf{\normalsize bunny} & \textbf{\normalsize spnza} & \textbf{\normalsize chsnt} & \textbf{\normalsize bath} & \textbf{\normalsize ref} & \textbf{\normalsize crnvl} \\ \hline
Tree Size(MB) & 0.2  & 0.5  & 12.2 & 22 & 25.5 & 104.2 & 37.1 & 37.3 \\ \hline
Depth     & 7 & 12 & 11 & 16 & 12 & 16 & 13 & 16 \\ \hline

Scene  &   \includegraphics[width=0.085\textwidth]{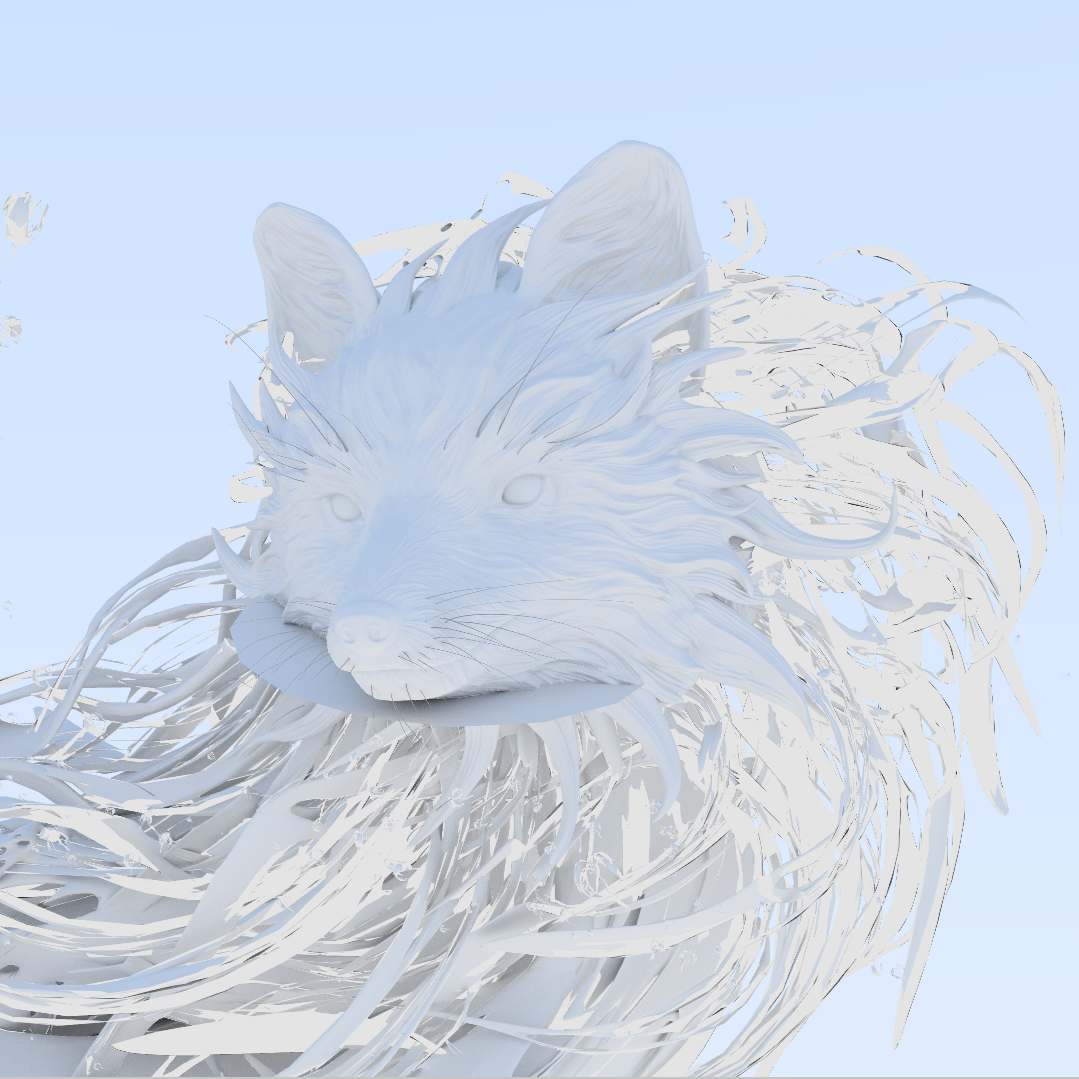} & \includegraphics[width=0.085\textwidth]{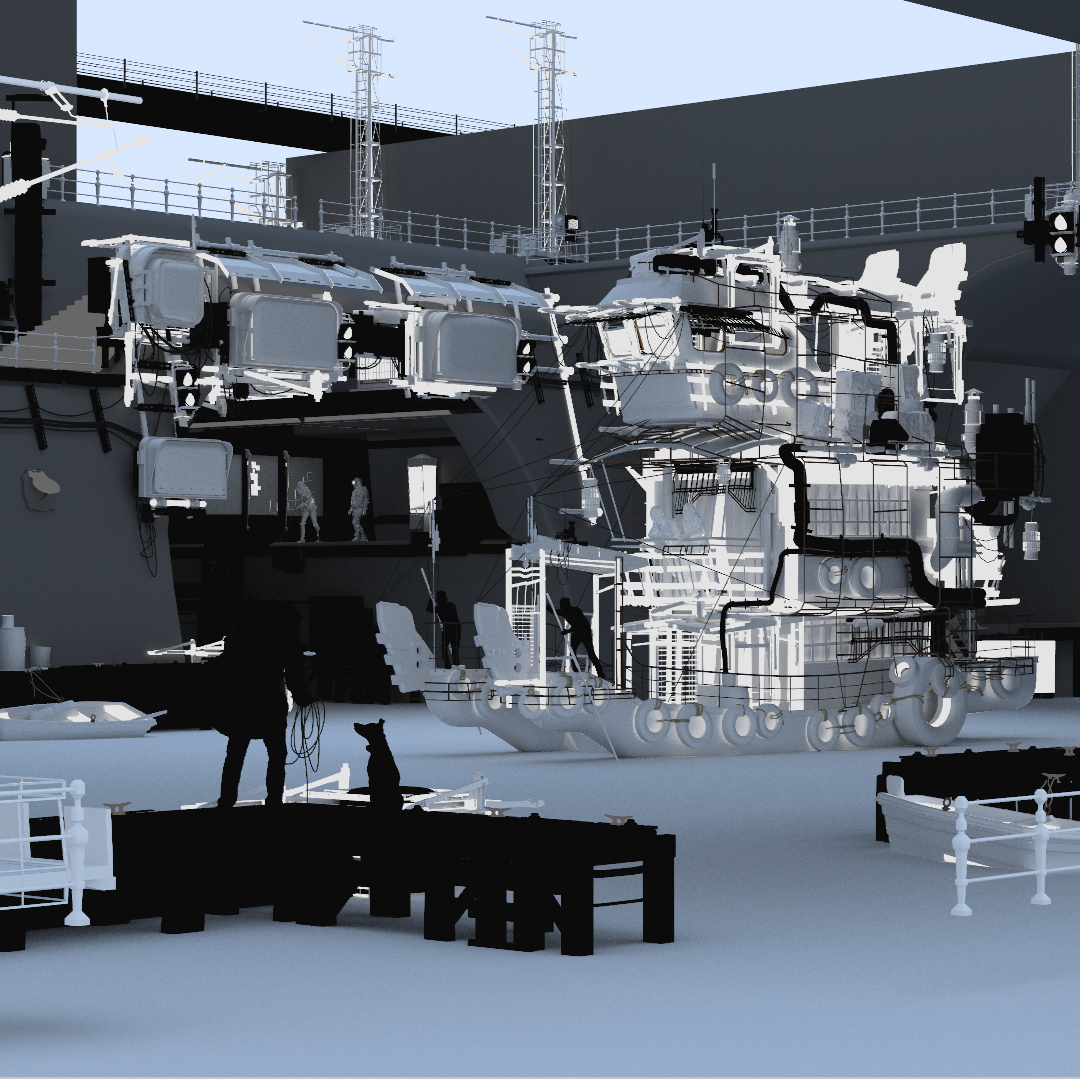} & \includegraphics[width=0.085\textwidth]{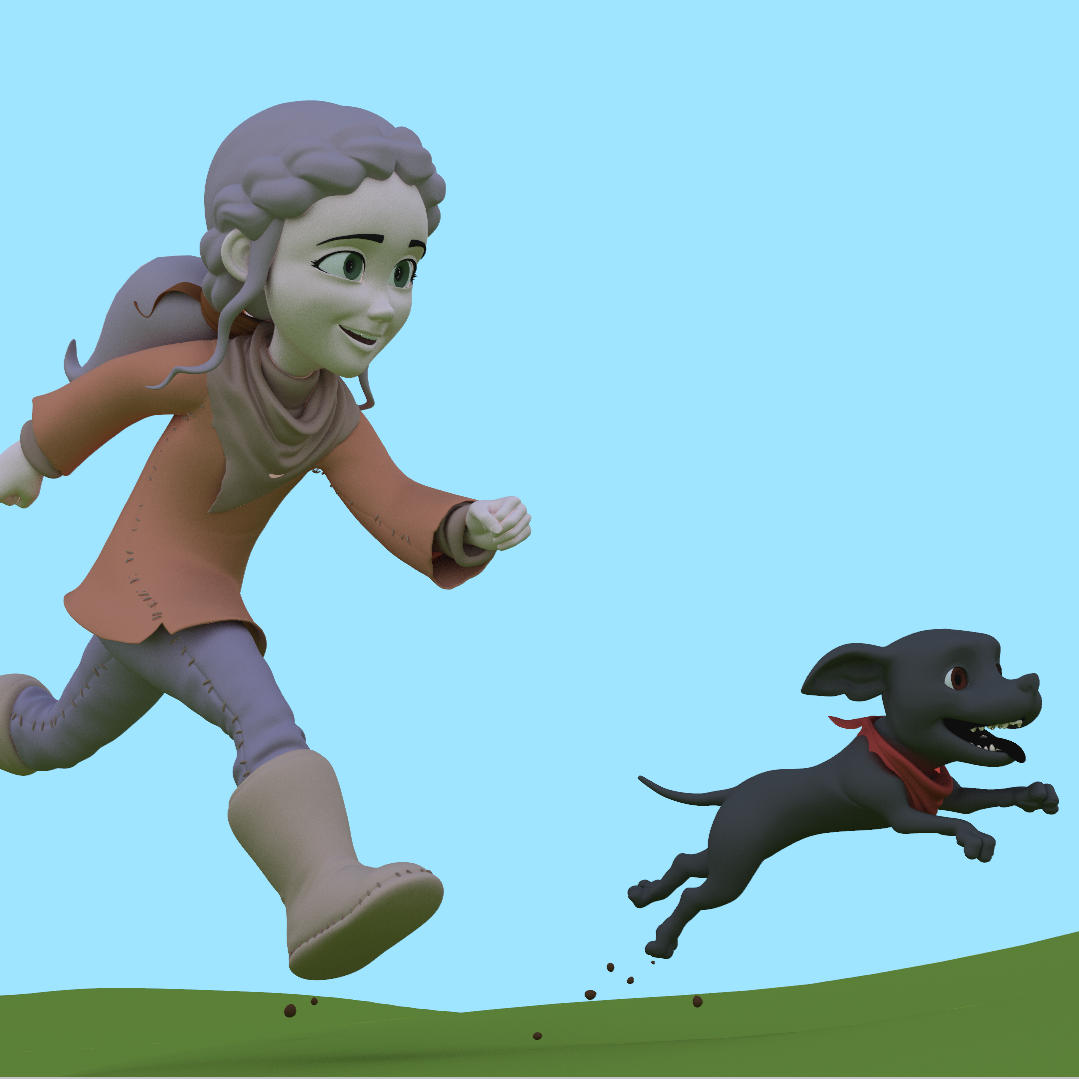} & \includegraphics[width=0.085\textwidth]{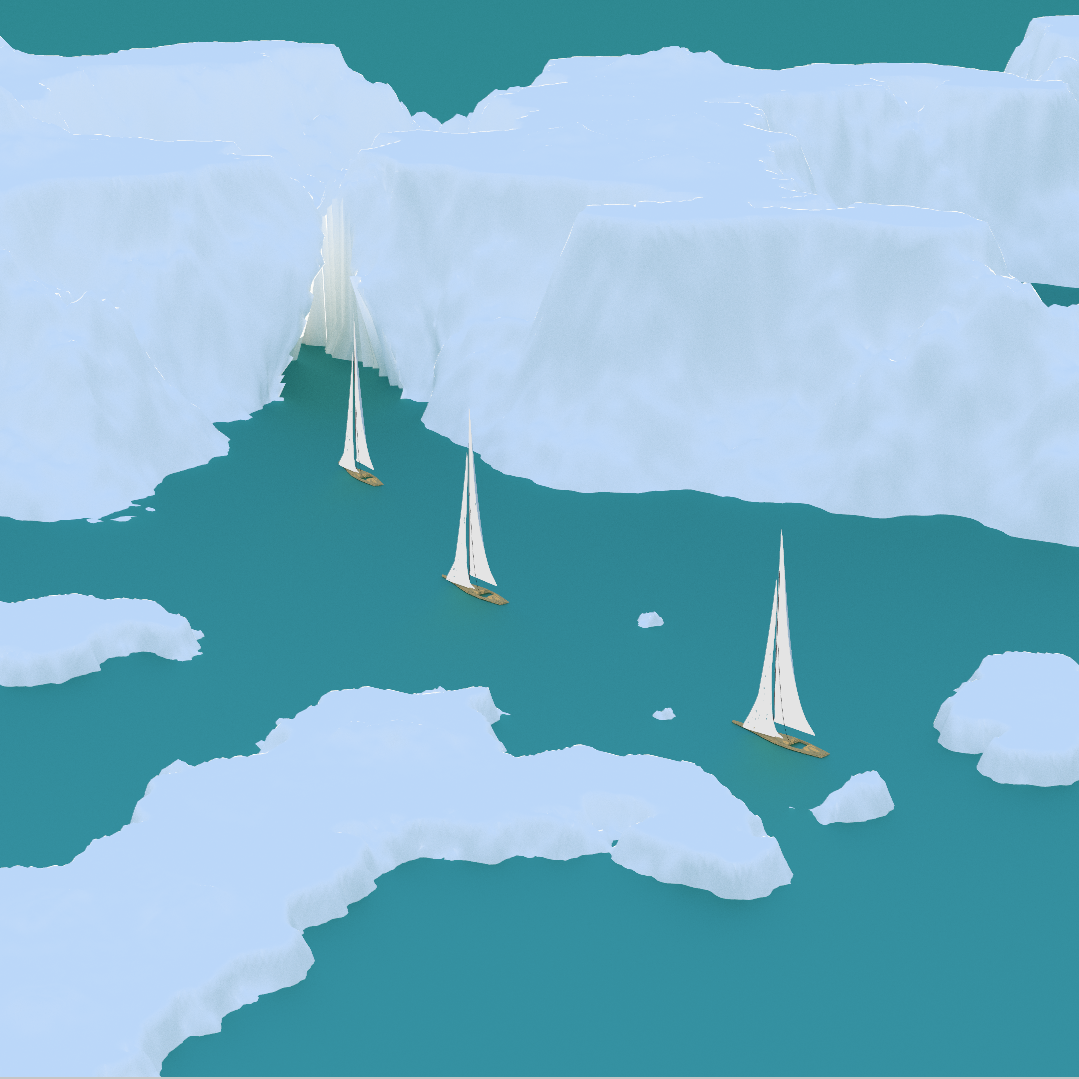} & \includegraphics[width=0.085\textwidth]{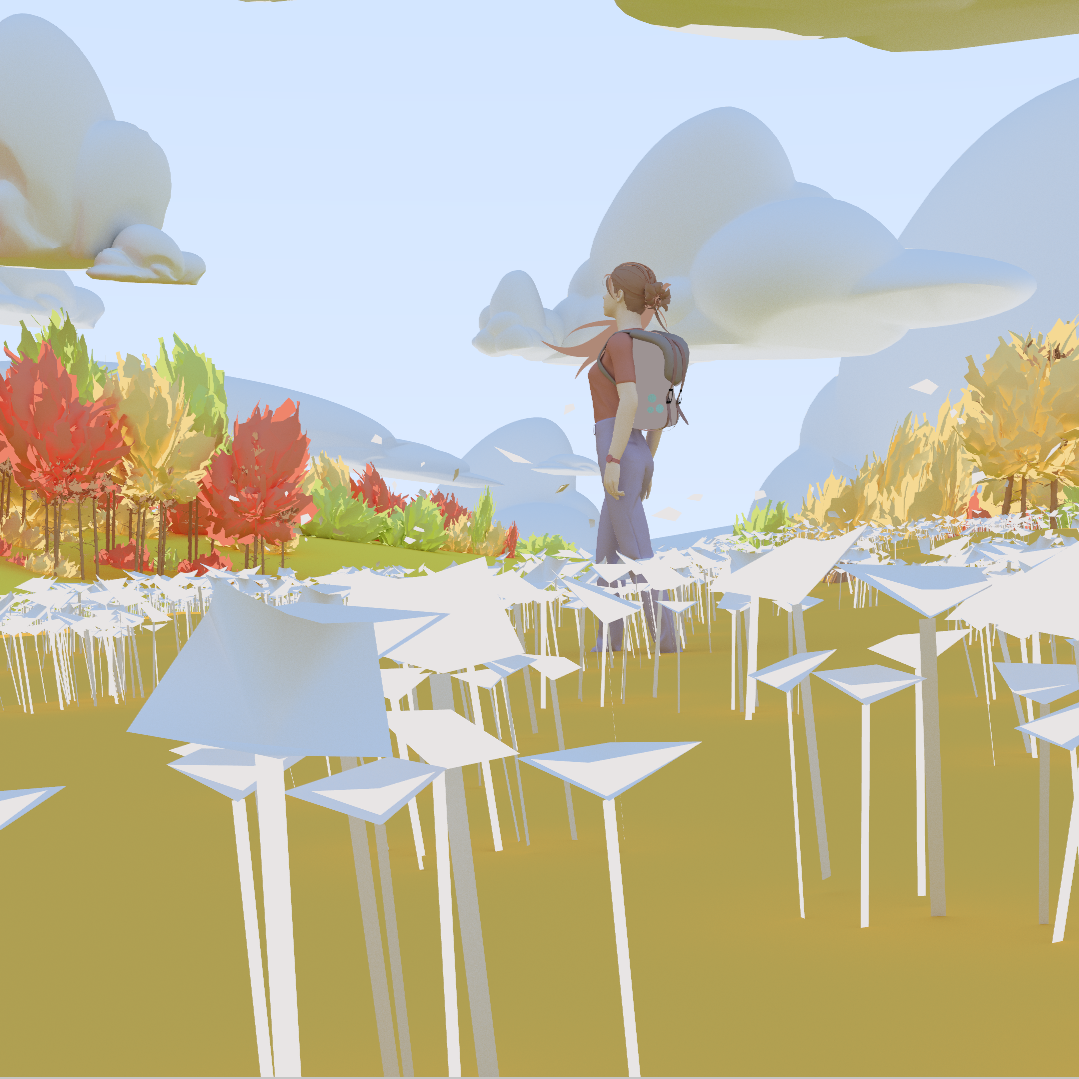} & \includegraphics[width=0.085\textwidth]{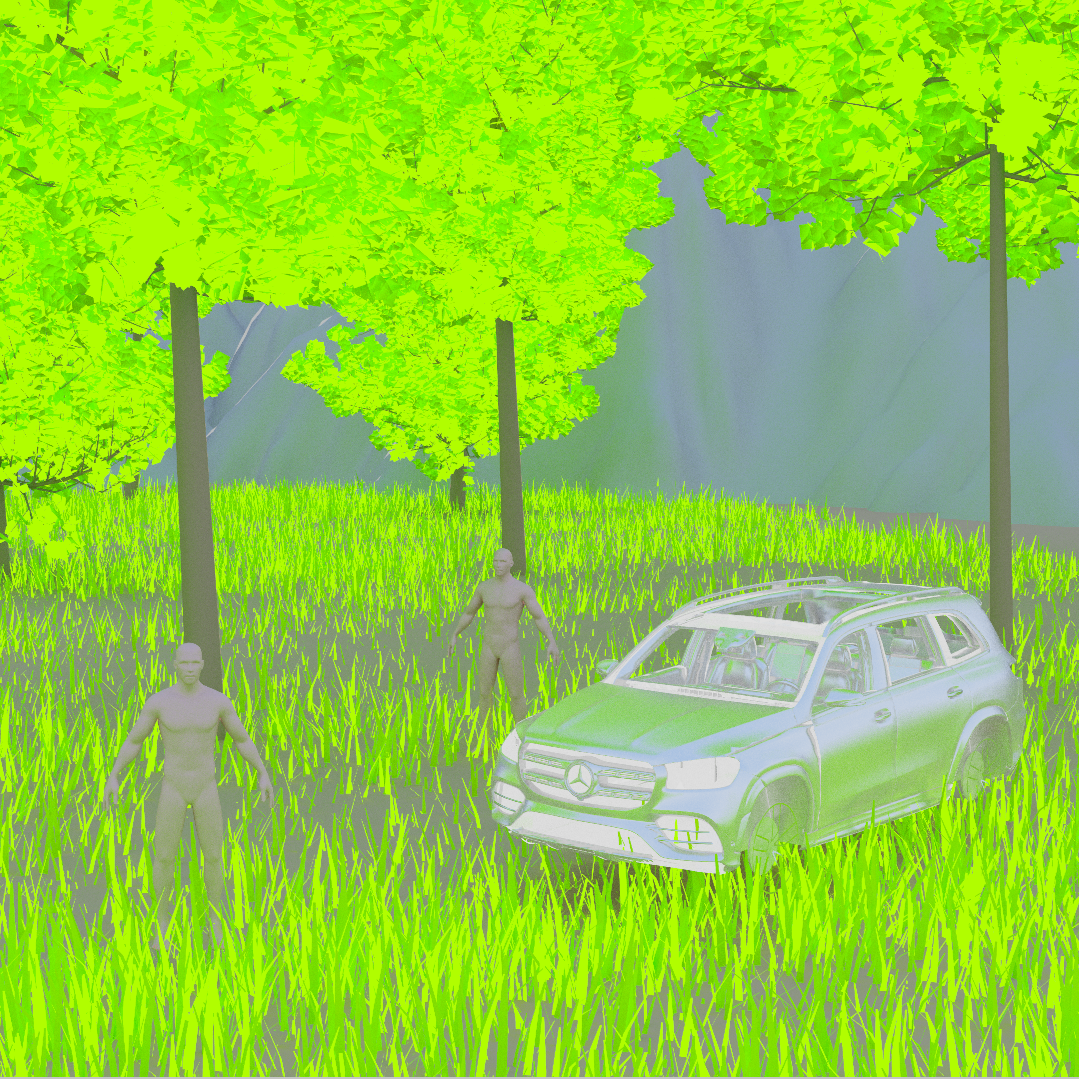} & \includegraphics[width=0.085\textwidth]{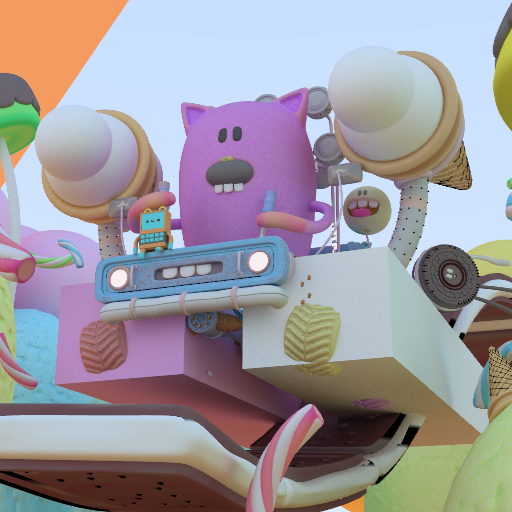} & \includegraphics[width=0.085\textwidth]{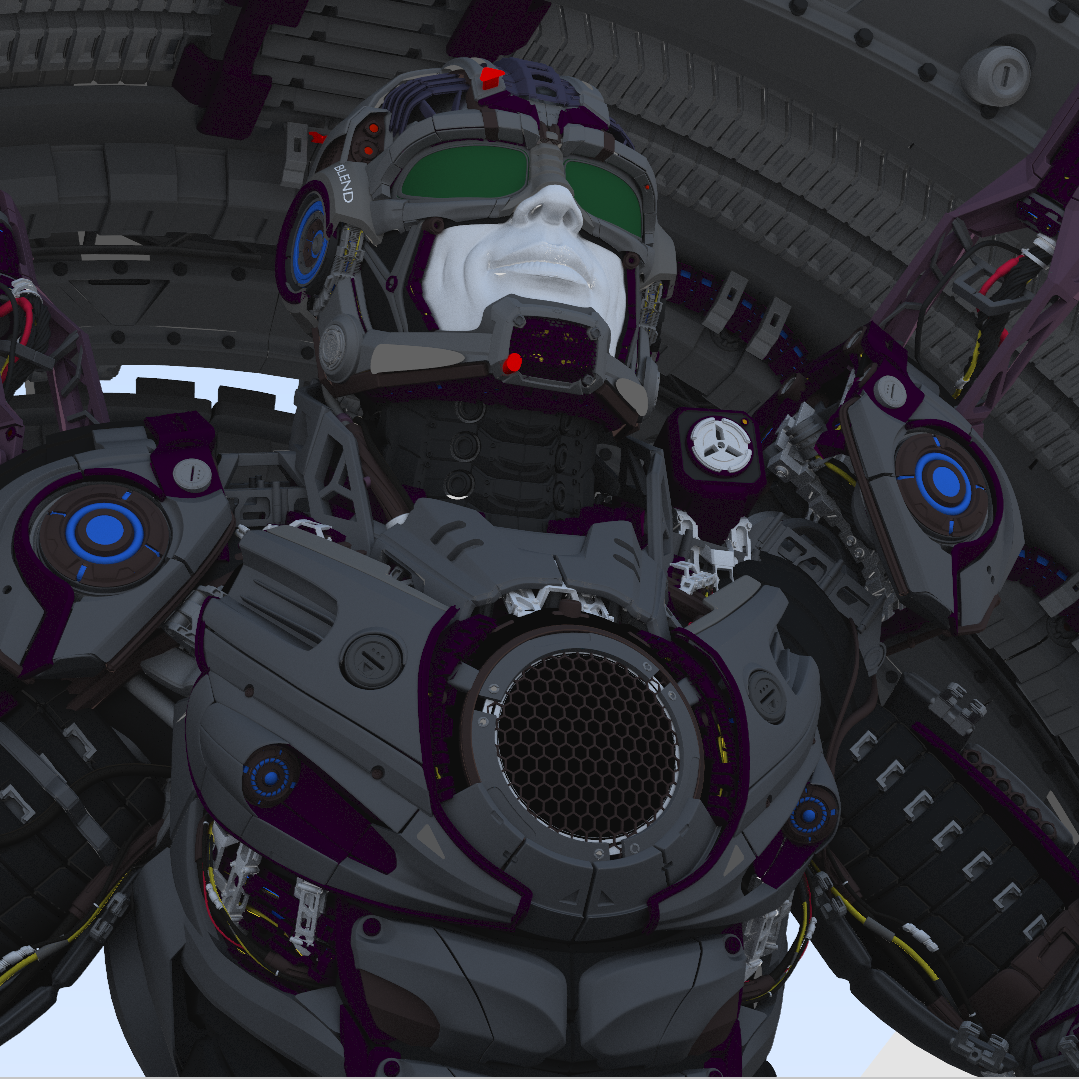}  
\\ \hline
Label     & \textbf{\normalsize fox} & \textbf{\normalsize party} & \textbf{\normalsize sprng} & \textbf{\normalsize lands} & \textbf{\normalsize frst} & \textbf{\normalsize park} & \textbf{\normalsize car} & \textbf{\normalsize robot} \\ \hline
Tree Size(MB) & 597.8  & 143.8  & 164.3 & 279.2 & 348.6 & 501.9 & 1,233.6 & 1,721.3 \\ \hline
Depth     & 15 & 14 & 14 & 12 & 14 & 14 & 16 & 18\\ \hline
\end{tabular}

\label{scene_table}
\end{table*}

We compare our TTP with the state-of-the-art prefetcher for ray tracing, the Treelet prefetcher\cite{treelet}. 
Both prefetchers are run at various resolutions using the same simulator configuration shown in Table \ref{config_table}.
For the Treelet prefetcher, we cloned and built the code available on Github\cite{noauthor_code_nodate} without any modifications. 
This repository has the necessary code that modifies the BVH tree to form treelets, perform treelet based traversal and prefetching.
We enable the following options to turn on the Treelet prefetcher: \textit{treelet\_based\_traversal}, \textit{treelet\_prefetch} and \textit{keep\_accepting\_warps}. 
\textbf{chsnt} scene consistently crashes at every resolution when Treelet prefetching is enabled, and therefore not included in the results.

\begin{scriptsize}
\begin{table}[]
\centering
\caption{Vulkan-sim Hardware Configuration}
\fontsize{7.6}{10}\selectfont
\begin{tabular}{{|p{0.45\columnwidth}|p{0.45\columnwidth}|}}
\hline
\# Streaming Multiprocessors(SM) & 8 \\
\hline
Max. TBs per SM & 16 \\
\hline
Warp Size & 32 \\
\hline
Instruction Cache & 128KB, 20 cycles \\
\hline
L1 Data Cache & 32KB, Fully assoc. LRU, 20 cycles, 256 MSHRs \\
\hline
L2 Cache & 512KB, 16-way assoc. LRU, 160 cycles, 768 MSHRs\\
\hline
Core, Interconnect, L2 Clock & 1365 MHz\\
\hline
Memory Clock & 3500 MHz\\
\hline
\# of RT Units per SM & 1\\
\hline
RT Unit Warp Buffer Size & 4\\
\hline
\end{tabular}
\label{config_table}
\end{table}
\end{scriptsize}

\section{Results} \label{resultsection}
\subsection{Overall Performance Evaluation} \label{overallperfsec}
\begin{figure*}[]
\centerline{\includegraphics[width=0.97\textwidth]{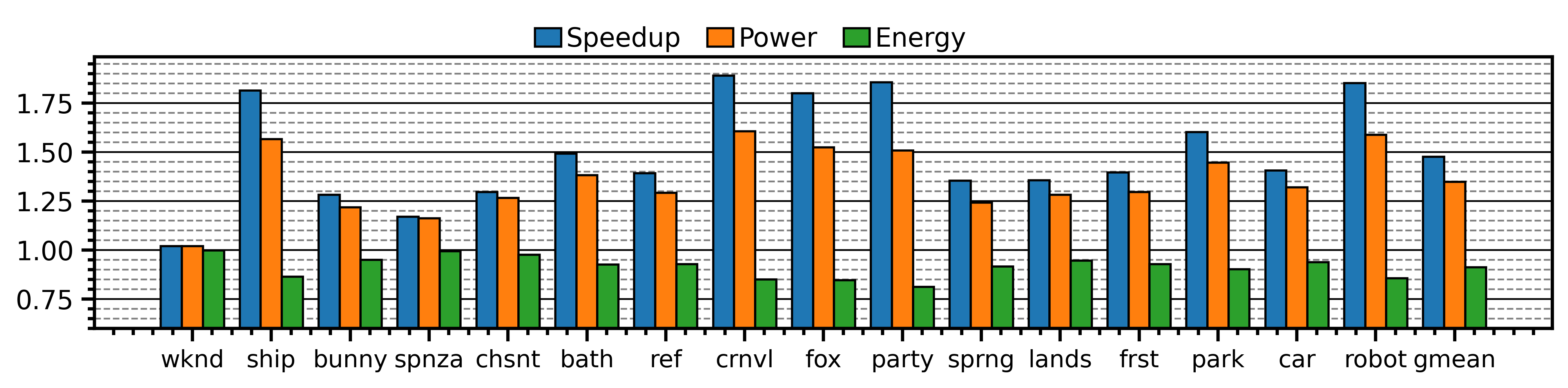}}
\caption{TTP speedup (higher the better), power and energy (lower the better) with DFS traversal, normalized to baseline.}
\label{dfsspeeduppowerenergy_pt}
\end{figure*}

We start with evaluating the overall performance of TTP.
Figure \ref{dfsspeeduppowerenergy_pt} shows the normalized speedup, energy and power over the baseline.
We see a geometric mean of 1.48x speedup (1.89x peak), 1.35x power and 0.91x energy consumption (i.e., 8.70\% energy savings).
Among the scenes, \textbf{wknd} is a very simple procedural scene featuring spheres and no triangles, offering limited room for improvement.
We observe that TTP yields the highest speedups in scenes with long pop streak misses (\textbf{ship}, \textbf{crnvl}, \textbf{fox}, \textbf{party}, \textbf{robot}), as shown before in Figure \ref{pop_miss_rates}.

\begin{figure}[]
\centerline{\includegraphics[width=0.48\textwidth]{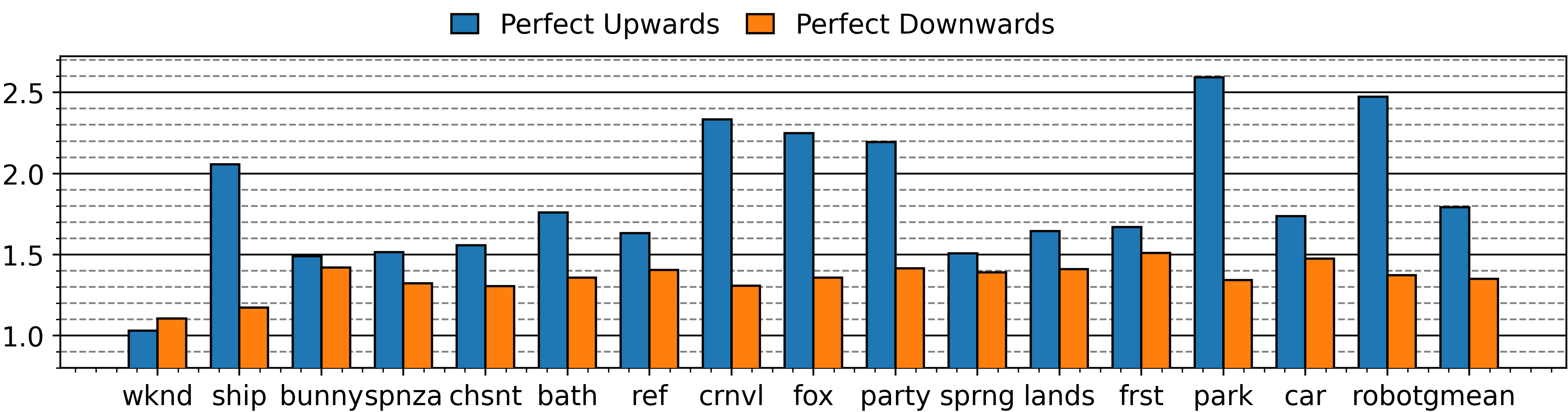}}
\caption{Normalized speedups with perfect upward and downward traversal.}
\label{perfectpref}
\end{figure}

In addition, we perform a limit study by simulating with perfect upward and perfect downward traversals. 
For perfect upward traversal, 2nd and later pops after a push always hit in the L1 cache, and for perfect downward traversal, 1st pops after a push always hit in the L1 cache.
Figure \ref{perfectpref} shows the results, where geometric means are 1.79x and 1.35x. 
These results correlate well with the characterization in Figure \ref{pop_miss_rates}.

\begin{figure}[]
\centerline{\includegraphics[width=0.48\textwidth]{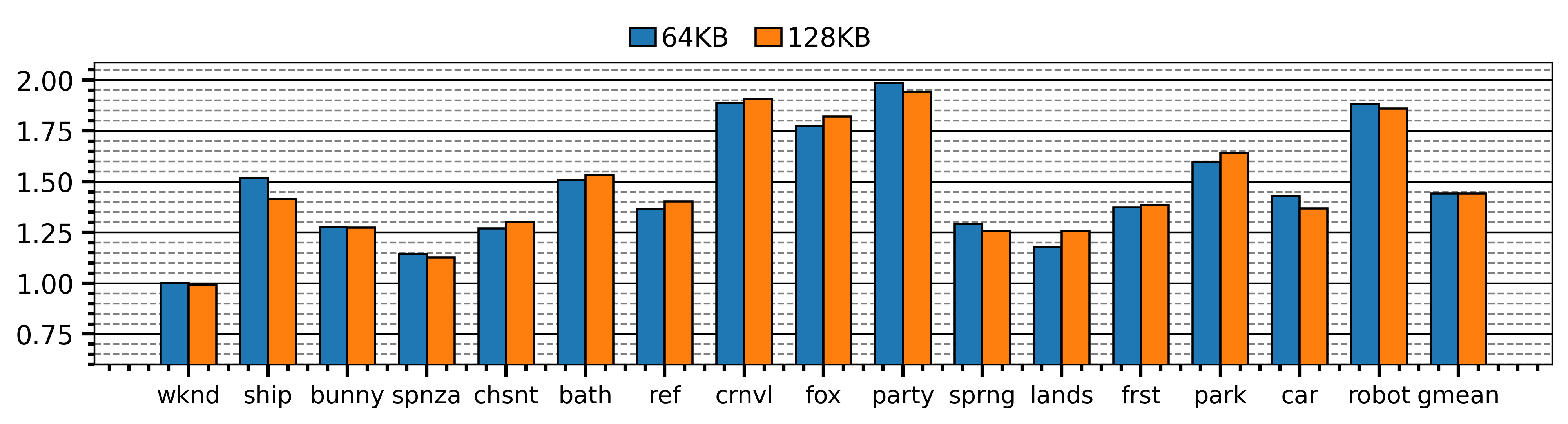}}
\caption{Normalized speedups with larger L1 data caches.}
\label{largedcache}
\end{figure}

TTP can also work with larger L1 data cache sizes. Figure \ref{largedcache} shows the normalized speedups (1.44x average for both) with cache sizes of 64KB and 128KB.

\begin{figure}[]
\centerline{\includegraphics[width=0.48\textwidth]{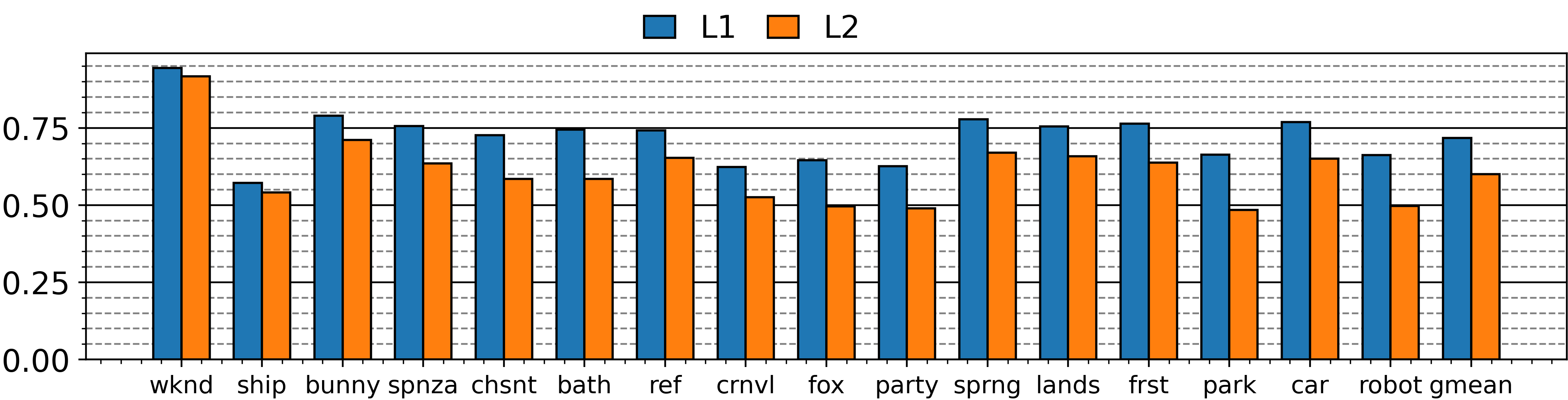}}
\caption{RT read MPKI with TTP normalized to baseline. Lower is better.}
\label{mpki}
\end{figure}

Figure \ref{mpki} shows how the L1 and L2 RT read misses-per-kilo-instruction (MPKI) rates change with TTP. 
Both L1 and L2 misses reduce consistently (28.28\% and 40.01\%, respectively) across all scenes, verifying the effectiveness of TTP.

\begin{figure}[]
\centerline{\includegraphics[width=0.48\textwidth]{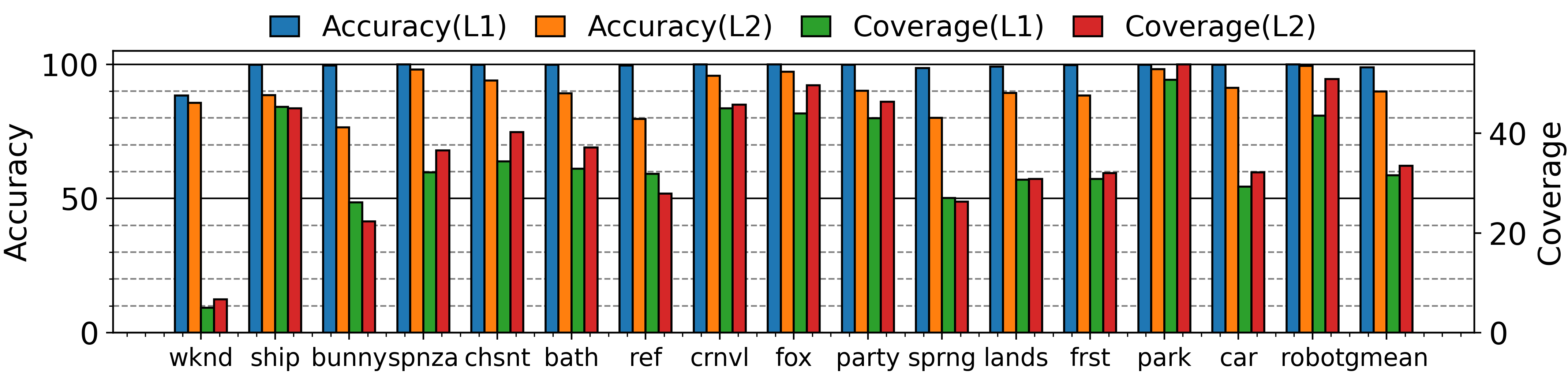}}
\caption{Prefetcher accuracy and coverage with DFS traversal. Higher is better.}
\label{dfs_accuracy}
\end{figure}

Figure \ref{dfs_accuracy} shows the accuracy and coverage of our TTP.
\textit{Accuracy} is the ratio of prefetched blocks that were accessed by a demand load. 
% \textcolor{red}{A prefetch is useful if it missed in cache and MSHRs, and the prefetched cache line accessed is by a demand read.}
Average accuracies are 98.92\% and 89.81\% for L1 and L2 respectively.
High levels of accuracy ensure that the cache is not polluted with unused data blocks.
The accuracy is not 100\% because in some rare cases, the prefetched data might end up being evicted before a demand
load accesses it.
\textit{Coverage} is the ratio of useful prefetches to all RT misses of baseline.
It indicates the proportion of misses that are prefetched and turned into hits.
Averages are 31.54\% and 33.46\% for L1 and L2 coverages, respectively.
Speedups are largely proportional to coverage. 
Higher coverage indicates that the BVH traversal for a particular scene exhibits the down-and-up traversal trend that our prefetcher performs well with. 

\begin{figure}[]
\centerline{\includegraphics[width=0.48\textwidth]{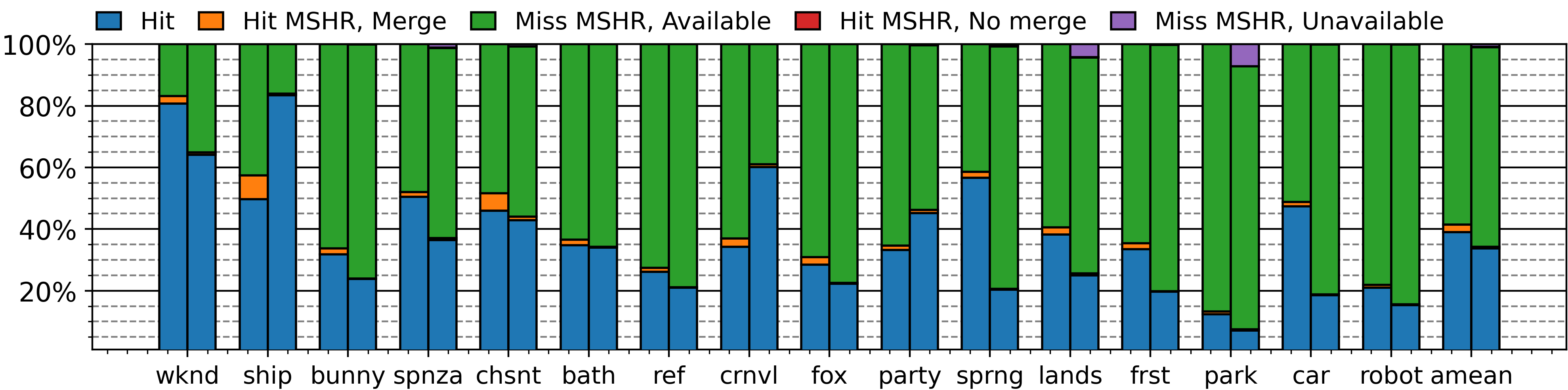}}
\caption{L1 and L2 cache responses to prefetch requests. For each scene, first column is L1, second column is L2.}
\label{cache_response_breakdown}
\end{figure}

To get a better understanding of the prefetcher efficiency, we look at the responses from caches. 
There are multiple possible outcomes of a cache access: (1) Hit, (2) hit in MSHR(Miss Status and Handling Register), or (3) miss in MSHR.
If there is an MSHR hit, the access request can be merged if a merge entry is available. 
If there is an MSHR miss, a new MSHR entry can be created if an empty entry is available.
We present the cache response breakdown for prefetch accesses for both L1 and L2 caches in Figure \ref{cache_response_breakdown}.
Note that the green bars, i.e. \textit{Miss MSHR, Available}, are the ideal cache response type for a prefetch request.
A prefetch request can only be useful if it misses in cache and MSHRs, and there is an empty MSHR entry available.
We define the prefetcher efficiency as the ratio of prefetch requests that miss in caches and MSHR, i.e. the green bars in Figure \ref{cache_response_breakdown}.
On average, the prefetcher efficiency is 58.56\% and 64.85\% for L1 and L2 respectively. Prefetches resulting in cache or MSHR hits are redundant but they do not pollute the cache. % prefetches occurring at the intra-thread, intra-warp, and inter-warp levels.
%However, most prefetch requests miss in caches and MSHRs. The total number of prefetch requests that miss in caches and MSHRs set an upper limit on the prefetcher accuracy.

\begin{figure}[]
\centerline{\includegraphics[width=0.48\textwidth]{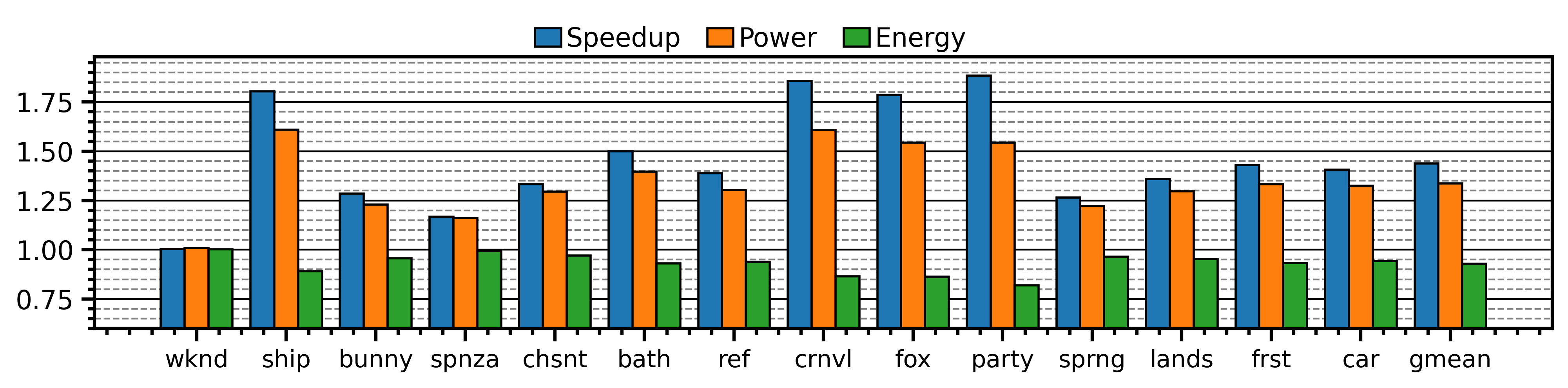}}
\caption{Normalized speedup, power and energy at 256x256 resolution. \textbf{park} and \textbf{robot} scenes time out after 72 hours at this resolution.}
\label{highres}
\end{figure}

By default, Lumibench uses 128x128 resolution for simulations. 
We increased the resolution to 256x256 to see the impact on TTP's performance.
Figure \ref{highres} shows the normalized speedups at 256x256.
Average speedup is 1.44x for 256x256, which is very close to the results at 128x128.

\begin{figure}[]
\centerline{\includegraphics[width=0.48\textwidth]{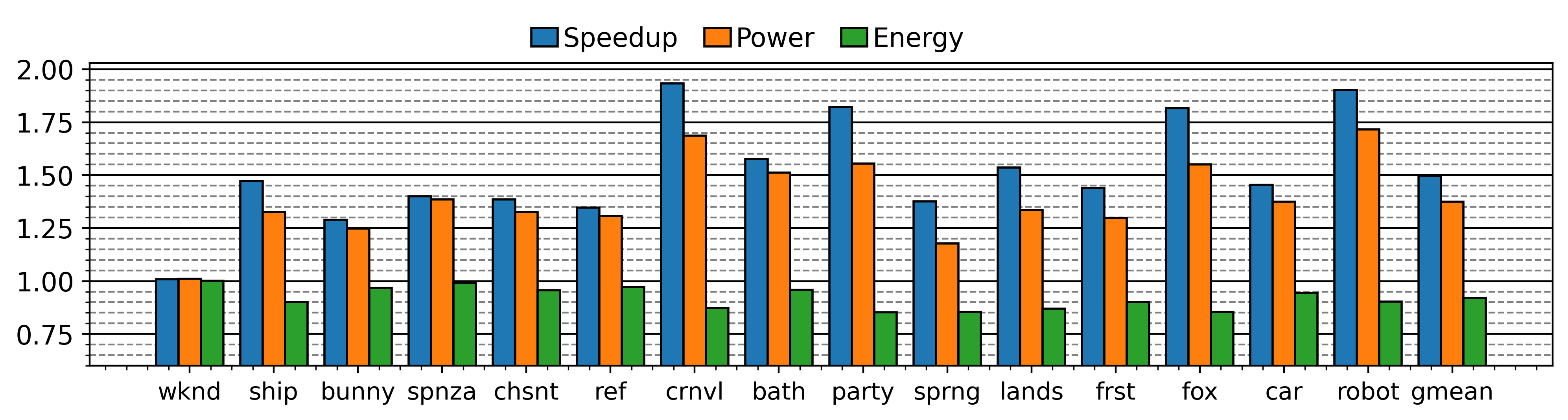}}
\caption{Normalized speedup (higher is better), power and energy (lower is better) with a larger GPU model.}
\label{desktop_speedup}
\end{figure}

To further establish TTP's robustness, we simulate it with a different hardware configuration that has 30SMs, 64KB L1 cache and 3MB L2 cache.
Figure \ref{desktop_speedup} shows the normalized speedups in this configuration.
Average speedup is 1.50x, and peak is 1.93x, which verify that TTP can work with different hardware setups.

\subsection{Prefetcher Arbitration and Aggressiveness} \label{agressec}
\begin{figure}[]
\centerline{\includegraphics[width=0.48\textwidth]{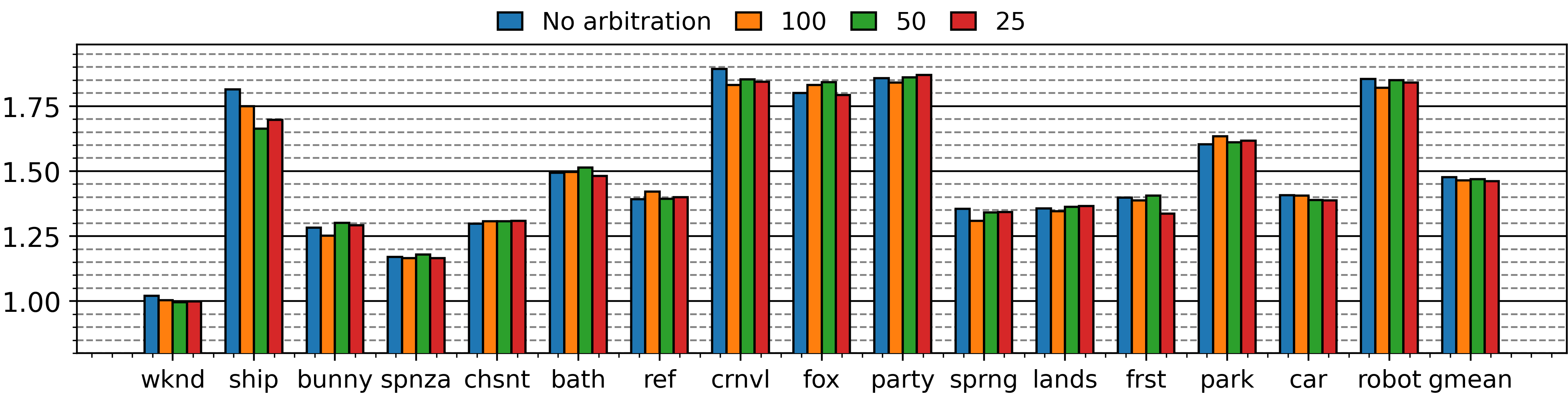}}
\caption{Speedups for different prefetch arbitration configurations, normalized to baseline. Higher is better}
\label{arbitration}
\end{figure}

We experiment with an arbitration scheme where prefetch requests are prioritized over demand requests if a fixed number of cycles have passed since the last prefetch request.
Results are shown in Figure \ref{arbitration}.
We evaluate thresholds of 25, 50, and 100 cycles. 
% Although some scenes, e.g. \textbf{ship}, \textbf{party}, \textbf{fox}, etc., prefer more aggressive prefetching, some such as \textbf{bath} and \textbf{spnza} prefer to prioritize demand loads.
On average, arbitration makes little to no difference.

\begin{figure}[]
\centerline{\includegraphics[width=0.48\textwidth]{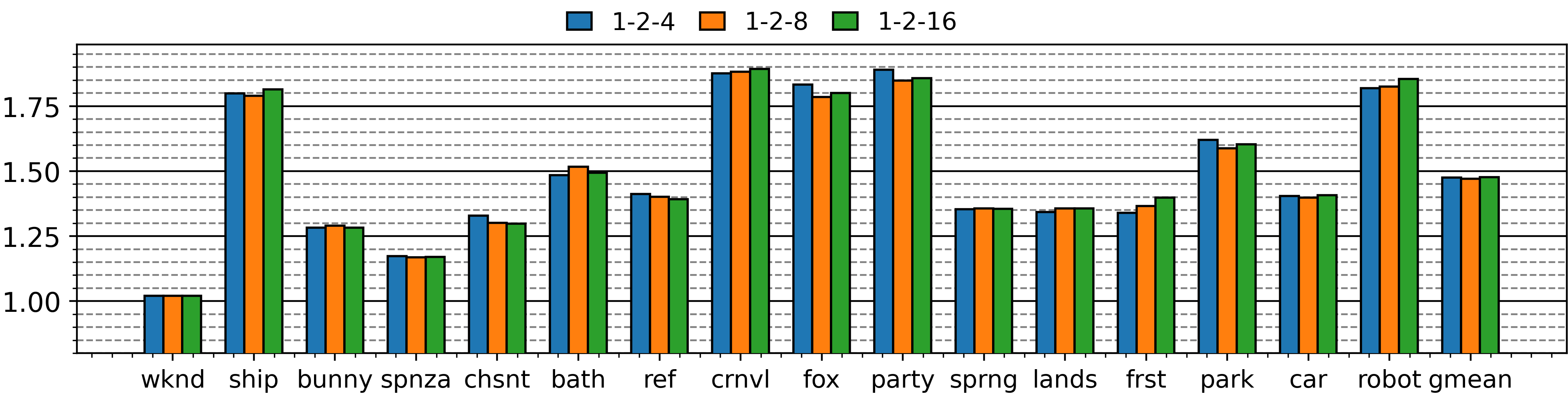}}
\caption{Speedups for different prefetch intensities, normalized to baseline. Higher is better.}
\label{dfs_aggressiveness}
\end{figure}

Figure \ref{dfs_aggressiveness} shows the speedups for different prefetch intensities of TTP. 
The numbers correspond to the number of prefetches that are generated in each state, as shown in the state machine diagram in Figure \ref{state_machine}.
All configurations have similar overall performance although certain scenes prefer higher/lower intensity. 
This shows that TTP is not sensitive to this parameter.

\subsection{TTP with BFS} \label{bfsressec}
\begin{figure}[]
\centerline{\includegraphics[width=0.48\textwidth]{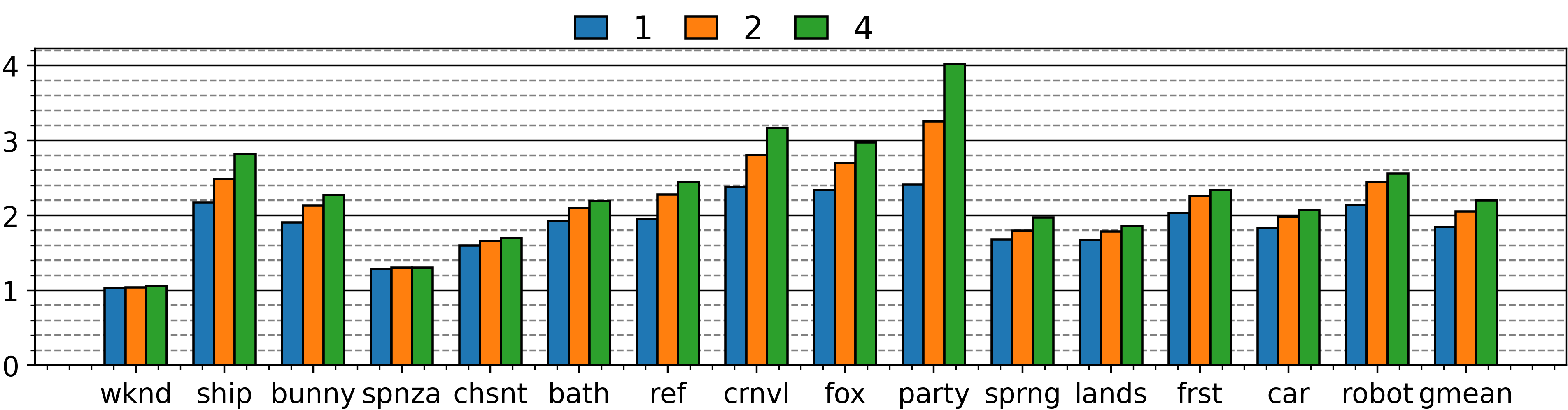}}
\caption{Normalized speedup for different prefetch distance values with BFS.}
\label{bfsspeedup_pt}
\end{figure}

Figure \ref{bfsspeedup_pt} shows the normalized speedup of TTP with BFS normalized to BFS without TTP.
We experiment with 3 different prefetch distance values: 1, 2 and 4.
TTP achieves an average of 1.85x, 2.05x, and 2.20x speedup for $N=1$, $2$, and $4$ respectively.
Longer prefetching distance is favorable, although the returns diminish quickly as $N$ is increased.
We use $N=4$ for the rest of simulations for BFS.

\begin{figure}[]
\centerline{\includegraphics[width=0.48\textwidth]{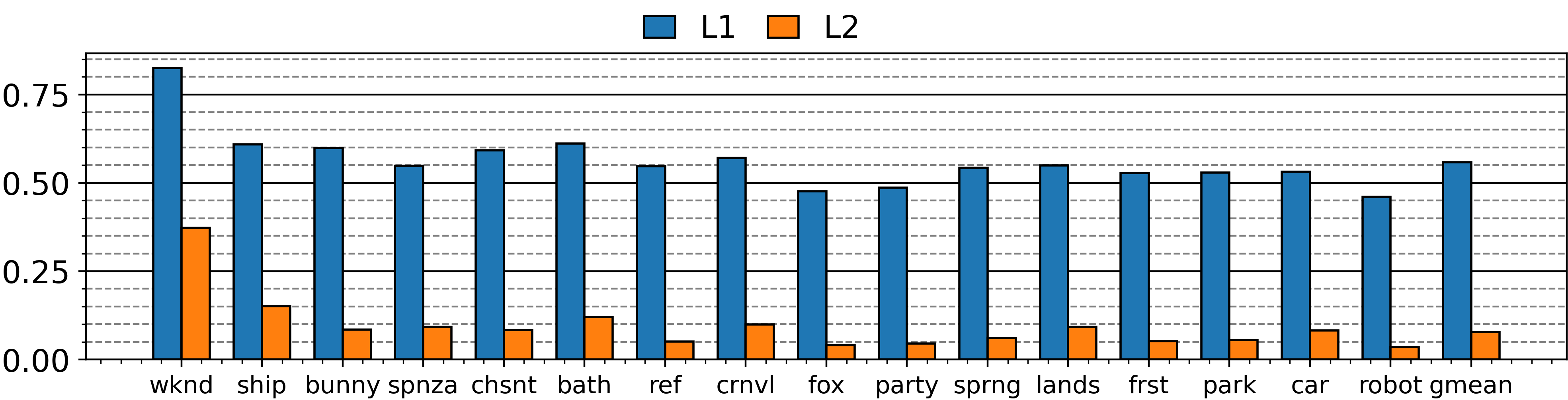}}
\caption{RT read MPKI with TTP normalized to baseline, with BFS \textit{N=4}. Lower is better.}
\label{bfs_mpki}
\end{figure}

Figure \ref{bfs_mpki} shows normalized RT read MPKI.
We note that both L1 and L2 misses are substantially reduced (44.10\% and 92.04\%, respectively), more so than the reductions in DFS.
As expected, prefetching in BFS is much more effective since it is predictable.

\begin{figure}[]
\centerline{\includegraphics[width=0.48\textwidth]{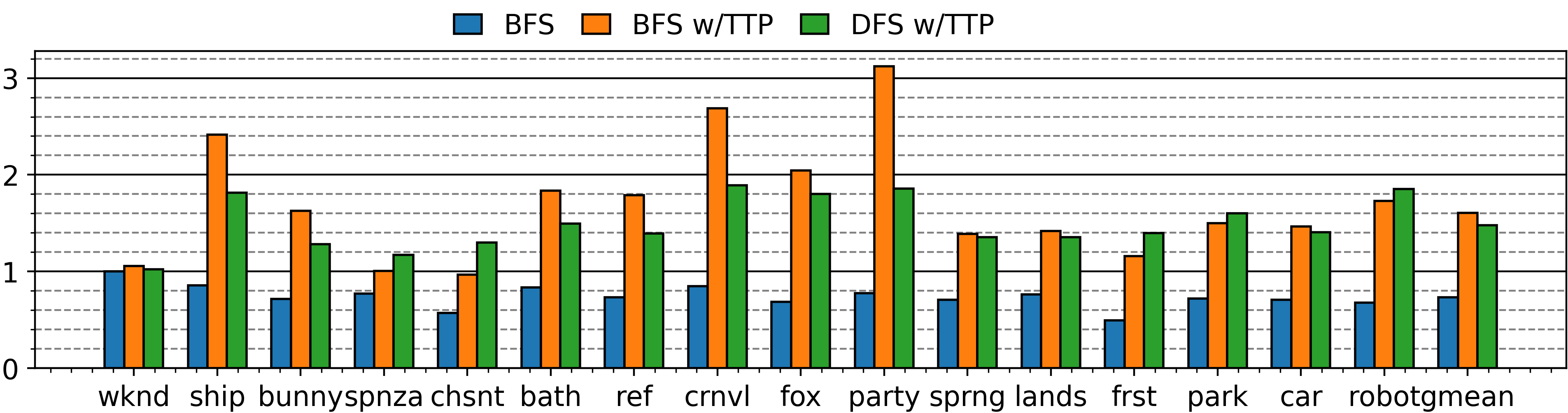}}
\caption{BFS and DFS speedups normalized to DFS without TTP. $N=4$ for BFS w/TTP. Higher is better.}
\label{bfsvsdfs}
\end{figure}

We also compare DFS and BFS, with and without TTP.
Figure \ref{bfsvsdfs} shows the speedups normalized to DFS.
When no prefetching is involved, BFS is always slower than DFS.
Interestingly, when TTP is enabled, BFS performs better than DFS on average, achieving 1.61x average speedup.
Although BFS accesses more nodes than DFS, it has better cache performance and therefore processes the nodes quicker than DFS.
The determining factor here is the \textit{average-nodes-per-ray} metric, which we presented in Table \ref{avg_nodes_table}.
DFS is faster in 5 scenes: \textbf{spnza}, \textbf{chsnt}, \textbf{frst}, \textbf{park} and \textbf{robot}.
These 5 scenes also have the highest average-nodes-per-ray difference between DFS and BFS, and consequently, BFS loses its edge over DFS.

\subsection{Comparison with Treelet Prefetcher} \label{treeletcompare}
\begin{figure}[]
\centerline{\includegraphics[width=0.48\textwidth]{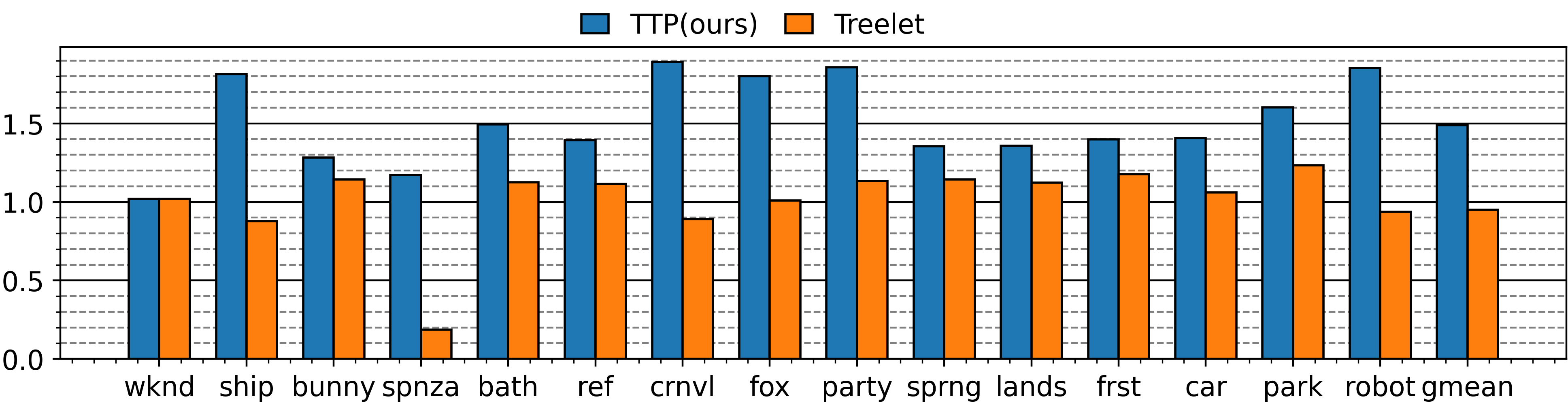}}
\caption{TTP and Treelet prefetcher speedups normalized to baseline, 128x128 path tracing. Higher is better.} 
\label{treeletvs_pt}
\end{figure}

Treelet prefetcher is a GPU hardware prefetcher targeting ray tracing\cite{treelet}. 
Treelet prefetcher requires the Treelet traversal algorithm, which involves pre-processing the BVH tree to divide it into treelets (sub-trees), and it directly replaces DFS. 
Once treelets are formed, the traversal is carried out in treelet granularity, i.e., all the nodes that are present within a treelet are traversed first before moving on to nodes in a different treelet.
A treelet is prefetched when its root node is read. 
This requires adding additional information in the BVH tree nodes to identify which treelet a BVH node belongs to during traversal. 
As such, this requires modifications to the traversal algorithm and the BVH tree organization. 
Another disadvantage of the Treelet traversal is that it may access more nodes than DFS does, because instead of traversing deeper into the tree to find a triangle, it prioritizes traversing treelets.
%To analyze the performance impacts of both prefetchers, we simulated them using the same configuration and shaders as ours at the same resolution.
%We use the vanilla code of the Treelet prefetcher from the Git repository\cite{noauthor_code_nodate}.
Figure \ref{treeletvs_pt} shows the speedups over baseline in path tracing for Treelet and TTP at 128x128 resolution.
TTP has better performance in all the scenes.
Treelet has 1.00x average speedup due to the performance loss in the scenes \textbf{ship}, \textbf{spnza}, \textbf{crnvl}, and \textbf{fox}.

\begin{figure}[]
\centerline{\includegraphics[width=0.48\textwidth]{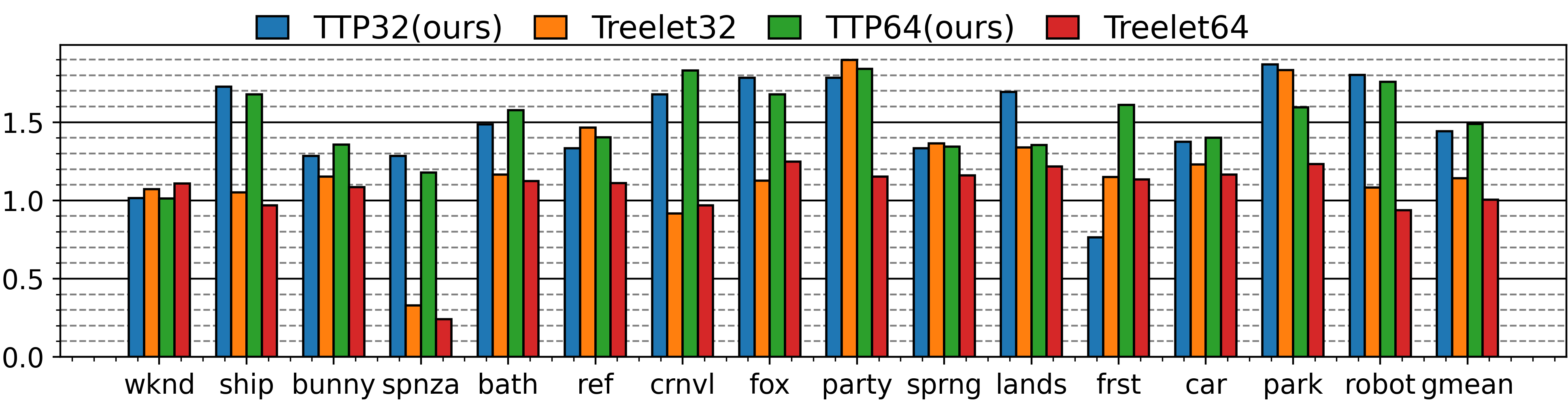}}
\caption{Speedups (higher is better) of the two prefetchers across different resolutions. For each scene, first two columns are 32x32, the last two columns are 64x64.} 
\label{treeletres}
\end{figure}

\begin{figure}[]
\centerline{\includegraphics[width=0.48\textwidth]{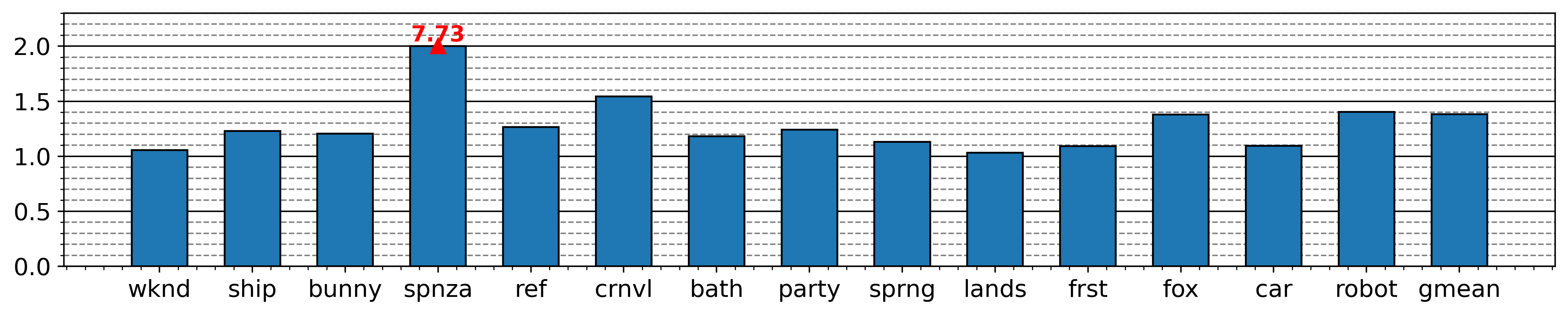}}
\caption{Total DRAM reads for Treelet prefetcher normalized to baseline. Lower is better.}
\label{treelet_dram_reads}
\end{figure}

To have a more comprehensive comparison, we simulate 32x32 and 64x64 frame resolutions. 
The results are presented in Figure \ref{treeletres}. 
When the frame resolution is changed to 32x32, Treelet prefetcher shows 1.14x speedup, close to the results reported in the previous work \cite{treelet}, whereas TTP achieves 1.44x speedup.
It should be noted that, at this low resolution, the SMs are under utilized due to the limited number of thread blocks. 
At 64x64, Treelet achieves 1.00x speedup, mainly due to performance regression in some scenes, whereas TTP maintains 1.49x speedup.
These results show that TTP outperforms Treelet and is more robust across different resolutions. 
Treelet prefetcher generates excessive prefetches which consume too much bandwidth. 
This especially becomes a problem as we increase the resolution from 32x32 to 128x128, which results in more thread blocks and therefore higher bandwidth utilization.
For example, in the \textbf{spnza} scene, average DRAM bandwidth usage goes up from 58\% in baseline to 80\% with Treelet at 128x128 resolution. 
It is likely that the excessive prefetches waste bandwidth and pollute the caches, leading to poor performance.
Figure \ref{treelet_dram_reads} shows the normalized number of total DRAM reads for Treelet prefetcher.
On average, the DRAM traffic increases by 1.38x.
For comparison, with our TTP, the total DRAM traffic does not change.

\subsection{Comparison with Park et al.\cite{park_node_2013}} \label{parketalcompare}
Park et al.\cite{park_node_2013} propose a prefetching scheme similar to TTP, the key difference being when and how often prefetches are triggered.
Park et al.’s prefetcher aims to overlap fixed-function math latency (i.e. leaf node intersection tests) with memory latency. 
For this approach to be effective, math latency should be roughly on par with the latency of a cache miss. 
This depends on the math hardware, BVH tree format(i.e. how many primitives are packed in each leaf node), and the memory subsystem performance. 
%camera
The BVH tree used in this work is built by the open-source Embree library\cite{embree}, which is what Vulkan-sim uses by default. The tree format has one primitive inside each leaf node, and the intersection latency is in the simulator is configured as 8 cycles for a leaf node, which is remarkably faster than the latency of a cache miss(100-300 cycles). 
If the BVH tree format packed more primitives in each leaf node, then the intersection test latency would potentially increase.
This could make it viable to overlap test latency with cache miss latency.
%In their work, Park et al. do not disclose the BVH tree format they used, nor the math latency, but it is implied that each leaf node contains multiple primitives.

\begin{figure}[]
\centerline{\includegraphics[width=0.48\textwidth]{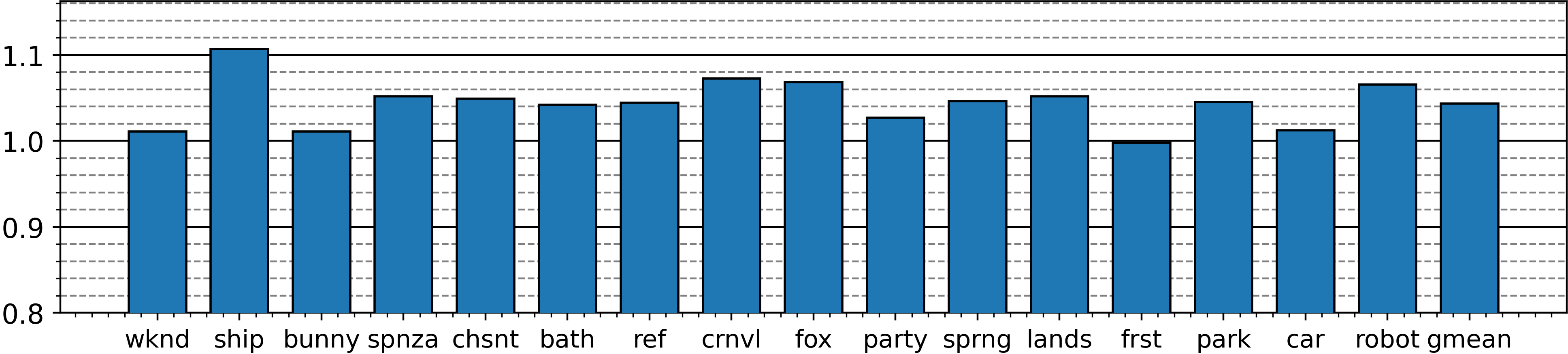}}
\caption{Normalized speedups with Park et al.'s prefetching strategy. Prefetches are generated during a leaf node intersection test to overlap test latency with memory fetch latency.}
\label{parketal}
\end{figure}

We model Park et al.’s prefetching strategy in Vulkan-sim to the best of our ability, and simulate using the same hardware configuration and resolution used for our TTP.
We generate prefetches during a leaf node intersection test, and  try to prefetch as many nodes as possible from
the stack during that interval. 
Figure \ref{parketal} shows the normalized speedups. 
We observed marginal speedups, averaging at 1.04x with a peak of 1.11x. 
The limited speedups are due to the timing of prefetches and how often the prefetcher is triggered.
The intersection latency is too short to hide a long memory request latency; and prefetching only happens after a leaf node is accessed. In contrast, TTP covers all upward traversals.

\subsection{Ambient Occlusion and Shadow Shaders}
Lumibench features two additional ray tracing workloads: Ambient Occlusion (AO) and Shadow (SH) shaders.
These shaders are used together with rasterization to improve visual quality, unlike path tracing which renders the frame by pure ray tracing.
Both shaders first trace a primary ray from user's position towards the scene and find the closest-hit point.
Then, AO shaders estimate the ambient light that reaches crevices of objects by tracing 4 randomly directed rays from the point of intersection.
SH shaders calculate shadows by tracing 2 rays from the hit point towards light sources.
Secondary rays, also called shadow rays, are any-hit rays, i.e., they terminate traversal on first hit.
In addition, they are more localized as they all share the same origin point (the intersection point of the primary ray).
Therefore, they are already relatively fast to execute, leaving small room for improvement, unlike path tracing.

\begin{figure}[]
\centerline{\includegraphics[width=0.48\textwidth]{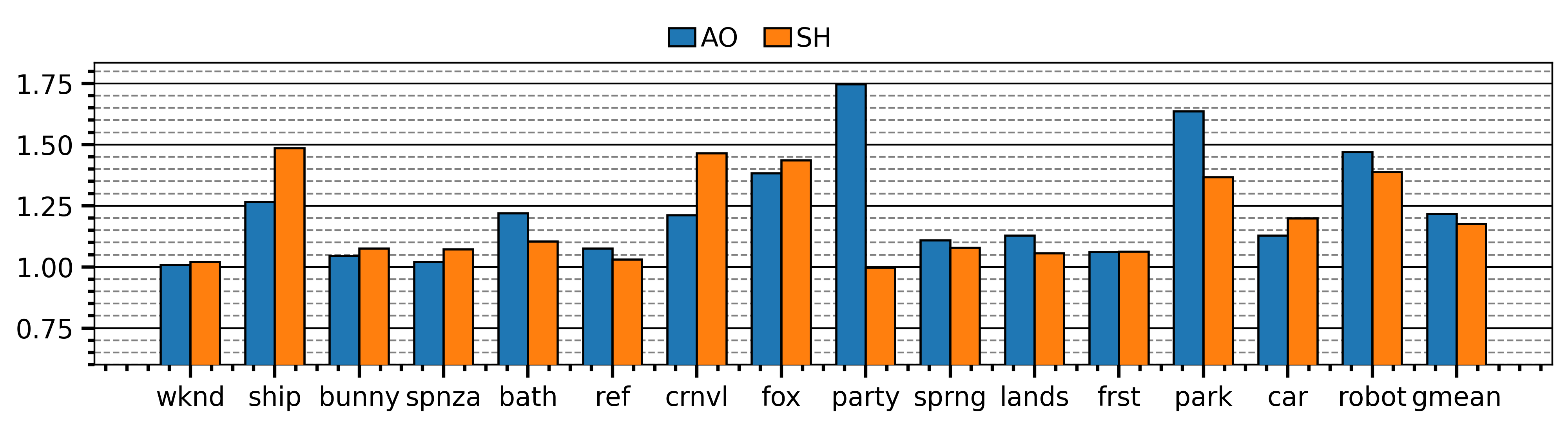}}
\caption{TTP speedups (higher is better) with AO and SH shaders, 128x128 resolution. \textbf{park} is 64x64. Lumibench does not support AO and SH shaders in \textbf{chsnt} scene.} 
\label{speedupaosh}
\end{figure}

We simulated TTP with AO and SH shaders, and achieved 1.22x and 1.18x speedup on average respectively, as shown in Figure \ref{speedupaosh}.
\textbf{chsnt} is not included in the results, because Lumibench does not support AO and SH shaders in this scene.

\subsection{Prefetcher Area Overhead}
To estimate TTP's area overhead, we implemented and synthesized the state-machine as described in Section \ref{prefsection} using FreePDK45\cite{stine_freepdk_2007}.
We assume the same hardware configuration as in Table \ref{config_table}, meaning there can be 4 warps in the RT unit, and each warp has 32 threads, therefore 128 state machines per SM.
The total number of cells needed to synthesize 128 state machines is 1117.
Each state machine has 2 bits of state, which requires 256 sequential cells.
The remaining 861 cells are used for the combinational logic.
On average, each state-machine requires $1117/128=8.7$ cells.
To put this into perspective, as shown in Figure \ref{gpu_diagram}, just the \textit{Ray Properties} field in the ray buffer would require $32*3*2=192$ bits of space per thread to store the ray origin and direction data.
Therefore, we conclude that the area overhead of TTP is negligible in comparison to the existing hardware.

\section{Related Work} \label{relatedworksection}
\subsection{Prefetchers}
%Talk about some gpu prefetchers, treelet etc.
Lee et al.\cite{lee_many-thread_2010} propose a GPGPU prefetcher that extends the CPU stride/stream prefetchers for many-core architectures. 
Their design generates prefetches on a warp granularity by detecting strides on a per-warp basis.
If multiple warps have the same stride, new warps can speculatively assume the same stride and generate prefetches.
Liu et al.\cite{liu_thread-aware_2016} use run-time information to dynamically adjust the prefetch amount per-thread and implement a filtering mechanism to discard harmful prefetches to avoid cache pollution.
Ainsworth et al.\cite{ainsworth_graph_2016} propose a graph prefetcher that targets BFS traversal.
As graphs can be more complex than BVH trees, their prefetcher design entails complexities that TTP does not require.
Specifically, their prefetcher snoops L1 accesses and determines which list(work, vertex, edge etc.) the address falls into. 
Then, it triggers a sequence of prefetches to fetch data from the lists, calculating a new address in between each prefetch.
By contrast, TTP only monitors the hardware traversal stack activity, which directly stores the node addresses without multiple levels of indirection that general purpose graphs require.
The Treelet prefetcher\cite{treelet} is a GPU prefetcher that targets ray tracing.
They propose a treelet based tree traversal algorithm, and prefetch at treelet granularity. 
Our proposed TTP achieves higher and more robust performance gains than Treelet as discussed in Section \ref{resultsection}.
%\textcolor{red}{The prefetcher that Park et al.\cite{park_node_2013} propose is the closest in spirit to TTP. The key difference is that their design only triggers prefetches when a leaf node is encountered, possibly to overlap the long leaf intersection test delay with the latency of the prefetch request. This means they only prefetch for a small subset of node reads, whereas TTP prefetches with every stack pop.}

\subsection{Accelerating Ray Tracing}
Hardware acceleration for ray tracing has been studied extensively\cite{aila_architecture_2010}\cite{aila_understanding_2009}\cite{rt_survey}. 
Using Vulkan-sim as a baseline, specialized prefetchers and cache structures have been proposed to mitigate the memory latency sensitivity of tree traversal\cite{treelet}\cite{lufei_cache}. 
Liu et al.\cite{lufei_cache} propose a dedicated cache structure that stores ray properties and intersected primitives for quick, speculative lookups to skip tree traversal.
Chou et al.\cite{chou_treelet_2025} propose a Treelet based RT unit design. 
Their solution is multifaceted, which involves virtual rays that increase the number of parallel rays in RT core, and dynamic shifting between traversal modes to accelerate traversal.
Overall, Chou et al. \cite{chou_treelet_2025} propose a more comprehensive and sophisticated solution, whereas TTP is a lightweight and low-overhead solution.

\section{Conclusion} \label{conclusionsection}
In this work, we propose a low-overhead, accurate hardware prefetcher, tree traversal prefetecher (TTP), for GPU accelerated ray tracing.
TTP leverages the readily available traversal stack in hardware to accurately prefetch BVH tree nodes in a timely manner.
For DFS based traversal, tree nodes are prefetched when consecutive stack pop operations happen, which indicate upward traversal.
We show that such upward traversals make up a large portion of all RT read misses.
As the per-thread traversal stacks store addresses of nodes that will be accessed next, this scheme requires no speculation of addresses, since the nodes in the stack will eventually be popped and read. TTP also readily supports BFS-based traversal, which is preferred for certain types of 3D scenes as well as general-purpose graph workloads.  

We evaluate TTP on Vulkan-sim 2.0, a cycle-level GPU architectural model, and show that it achieves up to 1.89x speedup with a geometric mean of 1.48x in path tracing workloads, while saving the overall energy by 8.70\%.
To estimate the hardware overhead of TTP, we synthesize the RTL for state machines, which consume negligible space compared to existing hardware.
We also compare our TTP with the state-of-the-art Treelet prefetcher specially designed for ray tracing.
We simulate both prefetchers at various resolutions and show that TTP outperforms the Treelet prefetcher with much less hardware and software overhead.

\section*{Acknowledgements}
We thank the anonymous reviewers for their valuable comments. The work is funded in part by NSF grants PHY-2325080 (with a subcontract to NC State University from Duke University), and OMA-2120757 (with a subcontract to NC State University from the University of Maryland).

\appendix
\section{Artifact Appendix}

%%%%%%%%%%%%%%%%%%%%%%%%%%%%%%%%%%%%%%%%%%%%%%%%%%%%%%%%%%%%%%%%%%%%%
\subsection{Abstract}

This artifact provides the source code for the modified Vulkan-sim which has newly added configuration options to enable Tree Traversal Prefetcher(TTP).
In addition, Python and shell scripts are provided to run all the necessary simulations and generate the plots and figures in this paper.
To streamline the artifact evaluation process, we prepared a docker image with all the software dependencies installed.
As simulations take a long time, we also included the raw simulation outputs that we generated and used for this paper.

\subsection{Artifact check-list (meta-information)}

{\small
\begin{itemize}
  \item {\bf Program:} Vulkan-sim, RayTracingInVulkan
  \item {\bf Compilation:} gcc/g++, ninja, meson, cmake, nvcc
  \item {\bf Run-time environment:} Ubuntu 20.04
  \item {\bf Hardware:} 32+GB RAM, 35+GB Disk space
  \item {\bf Metrics:} Number of cycles, average power consumption
  \item {\bf Output:} Vulkan-sim simulation outputs, figures.
  \item {\bf How much disk space required (approximately)?:} 30GB
  \item {\bf How much time is needed to prepare workflow (approximately)?:} About 1 hour
  \item {\bf How much time is needed to complete experiments (approximately)?:} 5 minutes to generate figures. One week for Vulkan-sim simulations, if ran in parallel
  \item {\bf Publicly available?:} Yes
  \item {\bf Code licenses (if publicly available)?:} Yes
  \item {\bf Archived (provide DOI)?:} \href{https://doi.org/10.5281/zenodo.19394324}{https://doi.org/10.5281/zenodo.19394324}
\end{itemize}
}

%%%%%%%%%%%%%%%%%%%%%%%%%%%%%%%%%%%%%%%%%%%%%%%%%%%%%%%%%%%%%%%%%%%%%
\subsection{Description}

\subsubsection{How to access}
We provide the Docker image which has everything required to run the simulations and generate the figures.
You can download it from Zenodo using the archived link.
\subsubsection{Hardware dependencies}
Only requirement is 32+GB of memory.
\subsubsection{Software dependencies}
Only a Docker installation is required. All software dependencies are installed in the Docker image.

%%%%%%%%%%%%%%%%%%%%%%%%%%%%%%%%%%%%%%%%%%%%%%%%%%%%%%%%%%%%%%%%%%%%%
\subsection{Installation}
Download the docker image from Zenodo, and start a container using the commands below,
\begin{verbatim}
$ docker load < ttp-isca2026-ae.tar.gz
$ docker run -it ttp-isca2026-ae:1.0 /bin/bash
\end{verbatim}

%%%%%%%%%%%%%%%%%%%%%%%%%%%%%%%%%%%%%%%%%%%%%%%%%%%%%%%%%%%%%%%%%%%%%
\subsection{Experiment workflow}
Inside the container, we provide a shell script \verb|ttp_simulations.sh| that has all the simulation commands needed to generate results. 
However, due to the large number of simulations, we do not recommend running the shell script directly, as it runs the simulations sequentially.
Instead, depending on the resources, simulations should be run in parallel.
The shell script simply serves as a reference for simulation commands.
Workflow for launching parallel jobs depends on the system being used, therefore we cannot provide a one-for-all script to launch parallel simulations.
To launch a simulation in the container,

\begin{verbatim}
$ cd /home/root/vulkan-sim-root
$ source init_vars.sh
$ cd RayTracingInVulkan/build/linux/bin
$ ./RayTracer --scene 20 --width 32 \
--height 32 --samples 1 > ship_pt.log
\end{verbatim}

We also provide all of the raw simulation logs and the Python scripts that we used to plot the figures in this paper.
To generate the figures, following command can be used inside the container,
\begin{verbatim}
$ python3 figure1.py
\end{verbatim}
This will generate \verb|fig1.png| using the simulation logs under \texttt{ttp\_raw\_simulation\_logs}. Other figures can be generated in a similar fashion.
%%%%%%%%%%%%%%%%%%%%%%%%%%%%%%%%%%%%%%%%%%%%%%%%%%%%%%%%%%%%%%%%%%%%%
\subsection{Evaluation and expected results}
Running the Python scripts will generate the figures using the existing simulation logs. 
To reproduce or replace the simulation logs, the simulation commands in the shell script can be used.

%To address the memory latency bottleneck of ray tracing, we propose a novel hardware prefetcher.
%Our prefetcher uses the existing traversal stack hardware to predict node accesses and prefetches them shortly before they are needed.
%It outperforms the state-of-the-art prefetcher with much less hardware overhead.
%We do not require a custom traversal algorithm; both DFS and BFS traversal algorithms can be used.
%Our simulations show the the Traversal Trend Prefetcher outperforms the state-of-the-art Treelet prefetcher in most cases.
%Overall, these techniques improve performance by up to 21.6\% and 7.8\%  on average, with negligible hardware overhead.

%\section*{Acknowledgements}
%This document is an updated version of HPCA 2022 and 2023, which, in
%turn, has been derived from two previous conferences
%\clearpage
%%%%%%% -- PAPER CONTENT ENDS -- %%%%%%%%

%%%%%%%%% -- BIB STYLE AND FILE -- %%%%%%%%
\bibliographystyle{IEEEtranS}
\bibliography{references}
%%%%%%%%%%%%%%%%%%%%%%%%%%%%%%%%%%%%

\end{document}